\RequirePackage{amsmath}
\RequirePackage{mathtools}
\documentclass[12pt]{iopart}
\usepackage{iopams}
\usepackage{mathtools, amsmath, amsfonts,amstext, amssymb}
\usepackage{graphicx, color}
\usepackage{ifthen}
\usepackage{tikz}
\usetikzlibrary{calc}
\usetikzlibrary{shapes}
\usetikzlibrary{arrows}

\renewcommand{\color}[1]{}

\newcommand{\Prob}{\mathcal{P}}
\newcommand{\gs}{\mathrm{gs}}
\newcommand{\RSB}{\texttt{RSB}}
\newcommand{\UNSAT}{\texttt{UNSAT}}
\newcommand{\dd}{\mathrm{d}}

\newcommand{\abs}[1]{\left\lvert#1\right\rvert}

\begin{document}

\title[Mean-field avalanches in jammed spheres]{Mean-field avalanches in jammed spheres}
\author{S Franz$^1$ and S Spigler$^1$}
\address{$^1$ LPTMS, CNRS, Univ.\ Paris-Sud, Universit\'e Paris-Saclay, 91405 Orsay, France}
\ead{stefano.spigler@lptms.u-psud.fr}

\begin{abstract}
Disordered systems are characterized by the existence of many sample-dependent
local energy minima, that cause a step-wise response when the system is perturbed.
In this article we use an approach based on elementary probabilistic methods to
compute the complete probability distribution of the jumps (static avalanches)
in the response of mean-field systems described by replica symmetry breaking; we
find a precise condition for having a power-law behavior in the distribution of
avalanches caused by small perturbations, and we show that our predictions are
in remarkable agreement both with previous results and with what is found in
simulations of three dimensional systems of soft-spheres, either at jamming or
at slightly higher densities.
\end{abstract}

\maketitle

\section{Introduction}

\begin{figure}[h!]
\centering
\includegraphics[width=0.9\textwidth]{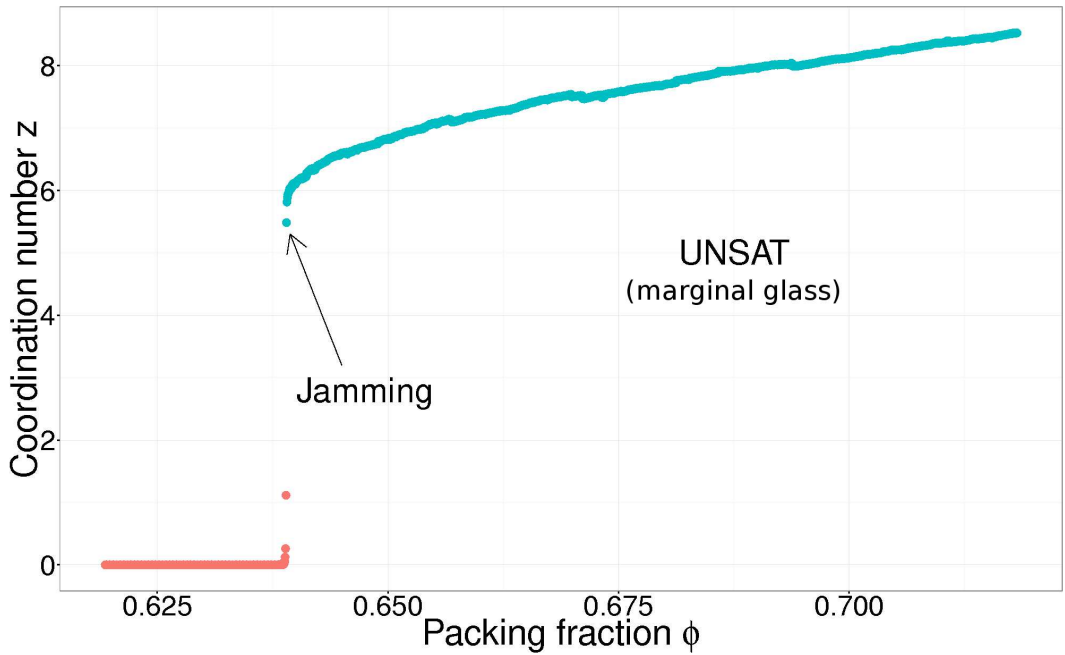}
\caption{Average coordination number (number of contacts per particle) as a function of the packing fraction, for
one specific 3-dimensional sample of 1000 soft spheres with harmonic repulsion.\label{fig:phiz}}
\end{figure}

Disordered systems have complex \emph{rugged} energy landscapes, with many
sample-dependent local minima that are deformed non-uniformly when the system is
slightly perturbed. {\color{blue}Upon increasing an external perturbation, a step-wise and sample-dependent
response is usually observed
\cite{yoshino2008stepwise,krzakala2002chaotic,young1984lack,young1982low,rizzo2006chaos,combe2000strain,sethna2001crackling,rosso2009avalanche,mueller2015marginal},
giving rise to jumps in several observables, called \emph{avalanches}.
An interesting feature is that the distribution of avalanches, averaged over
several samples, usually displays a \emph{power-law} behavior. Interestingly,
one often  finds the same power-law exponent using different perturbation
protocols, e.g.\ studying the response of the system under different dynamics
\cite{ledoussal2009frg,ledoussal2012equilibrium};} indeed, it has been
conjectured that in some disordered systems the various responses might lie in
the same universality class, regardless of the protocol~\cite{liu2009random}.
{\color{blue}When the protocol is such that the perturbation and the relaxation happen on separate
time scales (i.e.\ the relaxation is much faster than the perturbation),
the system always lies in its instantaneous ground state and the
avalanches are called \emph{static avalanches}; the numerical protocol that alternates
small perturbation steps and relaxation is usually referred to as the \emph{athermal
quasi-static protocol} (\texttt{AQS}). In this article we propose a derivation of the probability
distribution of static avalanches for a disordered system that can be
described by the \emph{replica symmetry breaking} (\RSB) framework; an alternative
derivation for the moments of such a distribution in the Sherrington-Kirkpatrick spin
glass model has been already presented in~\cite{ledoussal2012equilibrium}.
Our approach is based on elementary probabilistic methods exploiting the
Derrida-Ruelle probability cascades that characterize the Gibbs states, and in the case of the SK model it
leads to the same results found in~\cite{ledoussal2012equilibrium} via a
differential equations approach.\\ \\
Our motivation to re-examine this problem in a general framework comes
from the physics of soft spheres, in particular the jamming transition~\cite{ohern2003jamming,liu1998nonlinear,degiuli2014distribution,parisi2010mean,charbonneau2012universal},
where upon compression from a dilute phase, the system acquires mechanical stability. Soft spheres interact via a harmonic potential
when in contact, and do not interact when far away. When such a system has a low packing
fraction (ratio of the total volume of spheres to the total volume) it behaves as a
liquid: in particular, it does not provide any elasto-plastic response to external
forces and it has null shear-module. If the density is increased (while trying to minimize
the total energy of the system, i.e.\ trying to keep the spheres apart), at some point
the spheres will get in touch with each other: the packing fraction at which the average number of contacts
per particle is twice the space dimension (the \emph{isostatic} condition) is defined as the \emph{jamming} point, and
marks the onset of rigidity, since the number of contacts is sufficient to attain marginal
stability (this is the \emph{Maxwell's condition}). If the density is further increased
the average number of contacts increases as well, with a square-root dependence near the
jamming point, as shown in \Fref{fig:phiz}.}

\noindent{}{\color{blue}Soft sphere models have been recently analyzed and solved in the limit of
infinite dimensions~\cite{spheresI,spheresII,spheresIII}. Remarkably, the solution,
strongly based on \RSB{}, manages to predict properties of systems at jamming
that are apparently super-universal and that do not seem to depend
appreciably on the space dimension. In particular, the long-range contribution to the
distribution of small contact forces between particles at jamming follows a power-law
with a roughly constant exponent $\theta$ for any dimension. Moreover, recent
results~\cite{FranzParisiSevelevUrbaniZamponi} point out that two different critical
scaling solutions describe the jamming point and the \emph{jammed phase} (also called
\UNSAT{} phase, it is the phase found upon further compressing the system above jamming;
the name stands for ``unsatisfied'' and comes from the language of \emph{constrained
satisfaction problems}).
We are therefore interested in the implications of this scenario for static avalanches
induced by a shear-strain applied to the system. In this work we argue that
the exponent $\tau$ in the static avalanche distribution at jamming is related to the
small-force exponent $\theta$ via the relation $\tau = \frac{3+\theta}{2+\theta}$. This
is shown analytically in the infinite dimensional case, and numerical simulations suggest
that it is quite accurate in finite dimensions as well.
We also compare the statistics at jamming and in the \UNSAT{} phase: due to the different
scalings in the infinite dimensional models, the avalanche
exponents are different in the two cases, and there is a good agreement
between the finite and the infinite dimensional values, thus showing the non-trivial
presence of two different critical regimes in any dimension (at least for dimensions
larger than 3).\\ \\
The rest of the paper is organized as follows: in the next section we present a short
review on the structure of the states in a \RSB{} systems and how it can be represented
in terms of a stochastic process, known as Derrida-Ruelle cascade. In the section \emph{Approach}
we show how to use the Derrida-Ruelle process to compute the distribution of avalanches
in these systems.} In the section \emph{Asymptotic behavior} we discuss the result and the power-law
distribution of small jumps, that in section \emph{Simulations of systems of spheres}
will be compared with some numerical simulations for 3-dimensional systems of
soft spheres under shear. {\color{blue}Then, in \emph{Consequences on the elastic moduli}
we discuss the implications of the distribution of jumps that has been found at jamming
on the non-linear elastic moduli, comparing our results with~\cite{biroli2016breakdown}.}
In the \emph{Appendix} we present the detailed calculations.

\vspace{2em}
\section{Derrida-Ruelle cascades: a primer}

{\color{purple}
In this section we describe briefly, and for the scope of this paper, the picture
of the ergodicity breaking that emerges in the solution of mean-field glassy models.
In these systems, at sufficiently small temperature, the Gibbs measure is split in
ergodic components (\emph{pure states}) with nearly degenerate random free energies (i.e.\
with differences of order 1). To describe the organization of the space of these states
we introduce the notion of \emph{overlap} $q_{\alpha\beta}$ between two states $\alpha,\beta$: this is a co-distance
whose absolute value is normalized between 1 and 0 (for identical and maximally different states, respectively).
Different definitions are used for different systems: for a spin glass with $N$ spins (e.g.\ the
Sherrington-Kirkpatrick model) $q_{\alpha\beta} = \frac{1}{N}\sum_i s_i^\alpha s_i^\beta$, where $s_i^\alpha$
is the $i$-th spin in the state $\alpha$; for $N$ soft spheres, $q_{\alpha\beta} = \frac{1}{N}\sum_{ij=1}^N w(|\mathbf{x}_i^\alpha-\mathbf{x}_j^\beta|)$,
where $\mathbf{x}_i^\alpha$ is the position of the $i$-th particle in the state $\alpha$ and
$w(r)$ is a \emph{window} function that vanishes when $r$ is larger than some threshold.
The organization of the states is then \emph{ultrametric} with respect to the overlap, in the sense
that they are in a one-to-one relationship with the leaves of a rooted tree that
is generated via a \emph{Derrida-Ruelle cascade}~\cite{microstructure,mezard1985random,mezard1984replica,mezard2008spin,rsbnature,ruelle1987mathematical,aizenman2006mean,arguin2007spin}.\\ \\
The energy of any state in a given sample is the sum of an extensive, self-averaging
part (that is the same for all the states) and a term that is of order $\mathcal{O}(N^0)$. The Derrida-Ruelle cascade is a stochastic
branching process that describes the distribution of these non-extensive free energies in the different states.
It is usually described considering first a tree of finite depth $k$, and then taking the suitable limit for $k\rightarrow\infty$:
the process is fully characterized by $k$ pairs of increasing parameters $q_1<\cdots<q_k,\ x_1<\cdots<x_k$ that can
be thought of as a step-wise function $x(q,T)=x_i$ in $(q_i,q_{i+1})$;
the function $x(q,T)$ is known as the \emph{Parisi function}, and in the limit $k\rightarrow\infty$ it becomes continuous.
For the systems that we are dealing with, the function $x(q,T)$ is known from previous
works~\cite{spheresI,spheresII,spheresIII,FranzParisiSevelevUrbaniZamponi}, and is defined
as the solution to a variational problem; for small temperature $T$, $\beta x(q,T) \sim y(q) + \mathcal{O}(T)$
where $\beta$ is the inverse temperature and $y(q)$ is the zero-temperature limit of $\beta x(q,T)$, which
will be needed later to study the zero-temperature distribution of the states.

\begin{figure}[t]
\centering
\begin{tikzpicture}[scale=1.35]
  \draw [gray] (0,0) -- (-1.7,-1);
  \draw [gray] (0,0) -- (0.15,-1);
  \draw [gray] (0,0) -- (1.7,-1);

  \draw [gray] (-1.7,-1) -- (-2.6,-2);
  \draw [gray] (-1.7,-1) -- (-2,-2);
  \draw [gray] (-1.7,-1) -- (-1.4,-2);
  \draw [gray] (-1.7,-1) -- (-0.8,-2);

  \draw [gray] (0.15,-1) -- (-0.15,-2);
  \draw [gray] (0.15,-1) -- (0.45,-2);

  \draw [gray] (1.7,-1) -- (2.3,-2);
  \draw [gray] (1.7,-1) -- (1.7,-2);
  \draw [gray] (1.7,-1) -- (1.1,-2);

  \filldraw (0,0) circle(2pt);
  \filldraw (-1.7,-1) circle(2pt);
  \filldraw (0.15,-1) circle(2pt);
  \filldraw (1.7,-1) circle(2pt);
  \filldraw (-2.6,-2) circle(2pt);
  \filldraw (-2,-2) circle(2pt);
  \filldraw (-1.4,-2) circle(2pt);
  \filldraw (-0.8,-2) circle(2pt);
  \filldraw (-0.15,-2) circle(2pt);
  \filldraw (0.45,-2) circle(2pt);
  \filldraw (2.3,-2) circle(2pt);
  \filldraw (1.7,-2) circle(2pt);
  \filldraw (1.1,-2) circle(2pt);

  \node at (-2,-1) {$\alpha_1$};
  \node at (-0.15,-1) {$\alpha_2$};
  \node at (2,-1) {$\alpha_3$};

  \draw [thick] (3,0) -- (3.5,0); \node at (4,0) {$q_1$};
  \draw [thick] (3,-1) -- (3.5,-1); \node at (4,-1) {$q_2$};
  \draw [thick] (3,-2) -- (3.5,-2); \node at (4,-2) {$q_3$};

  \node at (0.45,0.75) {$\bar{F}_1$};
  \draw [->] (0.35,0.55) -- (0.1,0.15);

  \node at (-2.0,0) {$\bar{F}_{2,\alpha_1}$};
  \draw [->] (-2,-0.25) -- (-1.77,-0.85);

  \node at (1.3,-0.4) {$\beta x_1$};
  \node at (2.35,-1.4) {$\beta x_2$};

  \node at (-0.15,-2.35) {$F_\alpha$};
  \node at (1.7,-2.35) {$F_\beta$};
  \node at (2.3,-2.35) {$F_\gamma$};
\end{tikzpicture}
\caption{A realization of the Derrida-Ruelle cascade for a $k=2$ tree; a generic state $\alpha$ is explicitly shown.}
\label{fig:exampletree}
\end{figure}
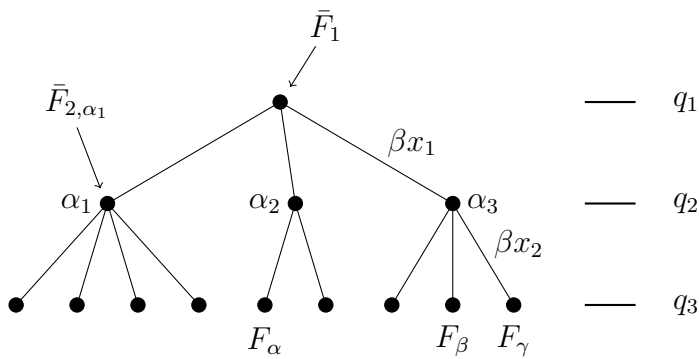\vspace{1.2em}

\noindent{}In order to describe the Derrida-Ruelle process, we start with an example of a $k=2$ tree,
as in~\Fref{fig:exampletree}. Starting from a reference free energy $\bar{F}_1$, that depends on the specific
sample, we generate the first level of the tree via a Poisson point process: the number of branches going
from the root node to nodes with free energy in $(\bar{F}_2, \bar{F}_2+\dd\bar{F}_2)$
is a Poisson variable with average $\exp({\beta x_1(\bar{F}_2-\bar{F}_1)})\dd\bar{F}_2$;
in this way we generate the nodes $\{\alpha_i\}$ with free-energies $\{\bar{F}_{2,\alpha_i}\}$.
Then, for each node $\alpha_i$ we generate sub-branches according to a new Poisson point process:
the number of branches going from $\alpha_i$ (with free energy $\bar{F}_{2,\alpha_i}$) to
nodes with free energy in $(F_3,F_3+\dd F_3)$ is again a Poisson variable, this time with
expected value $\exp(\beta x_2(F_3-\bar{F}_{2,\alpha_i}))\dd F_3$. The set of nodes in the last level
corresponds to the states of the system, and the free-energies of the states are precisely the values
generated via this process. We can also associate each overlap $q_i$ to the $i$-th level, as
shown in the figure: then, the overlap between two states (i.e.\ leaves) is simply
the value $q_i$ at the level of the closest common ancestor node; for instance, for the states $\alpha,\beta,\gamma$ in~\Fref{fig:exampletree},
$q_{\alpha\beta{}}=q_{\alpha\gamma{}}=q_1$ and $q_{\beta\gamma}=q_2$, while the self-overlaps
are all identical to $q_{\alpha\alpha{}}=q_3$ ($q_3\equiv q_{\mathrm{EA}}$ is known as
the Edwards-Anderson order parameter in spin glasses, and at zero temperature it becomes 1).
The general process for a tree with $k$ levels is very similar; the process is iterated
for all $k$ levels: the number of branches going from a node with free energy $\bar{F}_i$
at the $i$-th level to nodes with free energy in $(\bar{F}_{i+1},\bar{F}_{i+1}+\dd\bar{F}_{i+1})$
is a Poisson variable with average $\exp(\beta{} x_i(\bar{F}_{i+1}-\bar{F}_i))$. In the end
the non-extensive part of the free-energies of the states $\{F_\alpha{}\}$ are generated as the leaves of the tree.\\ \\}

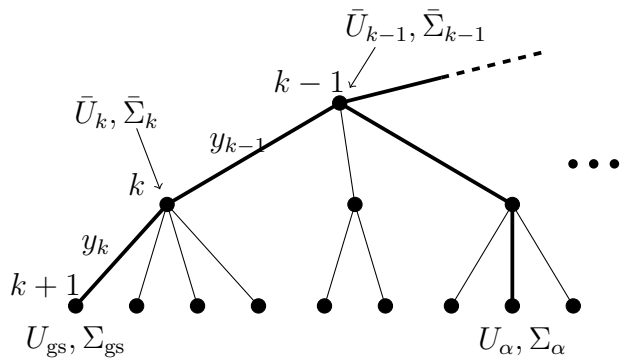
\begin{figure}[htb]
\centering
\begin{tikzpicture}[scale=1.35]

\draw [line width=0.5mm] (0,0) -- (1,0.25);
\draw [line width=0.5mm,dashed] (1,0.25) -- (2,0.5);
\node at (2.5,-0.6) {\Huge\bf...};

  \draw [line width=0.5mm] (0,0) -- (-1.7,-1);
  \draw [gray] (0,0) -- (0.15,-1);
  \draw [line width=0.5mm] (0,0) -- (1.7,-1);

  \draw [line width=0.5mm] (-1.7,-1) -- (-2.6,-2);
  \draw [gray] (-1.7,-1) -- (-2,-2);
  \draw [gray] (-1.7,-1) -- (-1.4,-2);
  \draw [gray] (-1.7,-1) -- (-0.8,-2);

  \draw [gray] (0.15,-1) -- (-0.15,-2);
  \draw [gray] (0.15,-1) -- (0.45,-2);

  \draw [gray] (1.7,-1) -- (2.3,-2);
  \draw [line width=0.5mm] (1.7,-1) -- (1.7,-2);
  \draw [gray] (1.7,-1) -- (1.1,-2);

  \filldraw (0,0) circle(2pt);
  \filldraw (-1.7,-1) circle(2pt);
  \filldraw (0.15,-1) circle(2pt);
  \filldraw (1.7,-1) circle(2pt);
  \filldraw (-2.6,-2) circle(2pt);
  \filldraw (-2,-2) circle(2pt);
  \filldraw (-1.4,-2) circle(2pt);
  \filldraw (-0.8,-2) circle(2pt);
  \filldraw (-0.15,-2) circle(2pt);
  \filldraw (0.45,-2) circle(2pt);
  \filldraw (2.3,-2) circle(2pt);
  \filldraw (1.7,-2) circle(2pt);
  \filldraw (1.1,-2) circle(2pt);

  \node at (-0.3,0.2) {$k-1$};
  \node at (-2,-0.8) {$k$};
  \node at (-2.9,-1.8) {$k+1$};

  \node at (0.75,0.75) {$\bar{U}_{k-1},\bar{\Sigma}_{k-1}$};
  \draw [->] (0.35,0.55) -- (0.1,0.15);

  \node at (-2.2,-0.1) {$\bar{U}_k,\bar{\Sigma}_k$};
  \draw [->] (-2,-0.25) -- (-1.77,-0.85);

  \node at (-1,-0.4) {$y_{k-1}$};
  \node at (-2.4,-1.4) {$y_k$};

  \node at (-2.6,-2.35) {$U_\gs,\Sigma_\gs$};
  \node at (1.8,-2.35) {$U_\alpha,\Sigma_\alpha$};
\end{tikzpicture}
\caption{A portion of an ultrametric tree for a $k$-\RSB{} system, the ground state and a generic state $\alpha$ are explicitly shown.}
\label{fig:ultrametrictree}
\end{figure}

{\color{blue}\noindent{}In order to study athermal systems of soft spheres under
shear-strain, we take the following considerations into account: at zero temperature
the free-energies of the states $F_\alpha$ become energies $U_\alpha$; the self-overlap
$q_{k+1}\equiv q_{\mathrm{EA}}$ becomes 1; the function $\beta x(q)$ is replaced by its zero-temperature limit, $y(q)$; we need the distribution of stresses in the states.
For systems of soft spheres (but the same holds e.g.\ for the Sherrington-Kirkpatrick model in a magnetic field),
energies and stresses are uncorrelated~\cite{emergencerigidity}, and it can be shown~\cite{microstructure}
that the stresses of the states can be generated with the following diffusion process on
the same ultrametric tree generated by the previous branching process: starting from
some reference stress $\bar\Sigma_1$ related to the sample, the stress
$\bar{\Sigma}_i$ of each node $i$ is Gaussian distributed with average $\bar{\Sigma}_{i-1}$ (its direct ancestor)
and variance proportional to $q_i-q_{i-1}$. Again, we continue the process until we reach the states, located on the leaves of
the tree. A portion of a ultrametric tree with $k$ levels is shown in \Fref{fig:ultrametrictree}.
In the end we want to take the continuous limit: without loss of generality we can
assume that $q_i=i\Delta q$, such that in the limit $k\rightarrow\infty$ we have
$\Delta q\rightarrow 0$, $k\Delta q\rightarrow 1$, and $y_i\rightarrow y(q)$.}

\vspace{2em}
\section{Approach}

Let us consider a system with a Hamiltonian $\mathcal{H}$ with $\mathcal{O}(N)$ degrees
of freedom (e.g. $N$ particles); we apply a small perturbation $\delta\gamma \ll 1$
that modifies the energy as $\mathcal{H}^\prime = \mathcal{H} - \frac{\delta\gamma}{\sqrt{N}} \Sigma$,
where $\Sigma$ is the suitable variable conjugated to the field. We will think
of $\delta\gamma$ and $\Sigma$ as the shear-strain and shear-stress for systems
of spheres, even though the results are general and apply also, for instance, to spin glasses
perturbed by a magnetic field.
The scheme of our approach is as follows: first, we need the distribution of
energies $\{U_i\}$ and stresses $\{\Sigma_i\}$ in the \emph{states} $\{\alpha_i\}$, that are
the local minima of $\mathcal{H}$; then, {\color{blue}since we are interested in the \emph{static} avalanches,
we need the distribution of the energy $U_{\beta}$ and stress $\Sigma_{\beta}$
of the new ground state, that is the state $\beta$ that minimizes the total energy
$U_{\beta}-\frac{\delta\gamma}{\sqrt{N}}\Sigma_{\beta}$ among all the states, for a given strain $\delta\gamma$ (notice that we will be dealing with the non-extensive part of the free energy only --- i.e.\ the one distributed according to the Derrida-Ruelle cascade, --- since the extensive term is the same for each state and needs not be taken into account).
Finally we will study the distribution of the difference $E_{\beta}-E_{\gs}$
($E_{\gs}$ being the energy of the unperturbed ground state)}
to check whether it displays a power-law behavior for small jumps at a fixed strain.
The distribution of this state $\beta$ can be written as a marginal distribution of the
joint probability of the states (we are integrating out all the other states, with the
constraint that they must have an energy higher than the new ground state $\beta$):

\begin{multline}
\Prob_{\min}[U_\beta,\Sigma_\beta, \beta | \delta\gamma] =\\= \int\left[\prod_{\alpha\neq\beta}\dd U_\alpha\dd \Sigma_\alpha \, \theta(U_\alpha - \frac{\delta\gamma}{\sqrt{N}}\Sigma_\alpha - U_\beta + \frac{\delta\gamma}{\sqrt{N}}\Sigma_\beta)\right] \Prob_\mathrm{states}[\{U_\alpha,\Sigma_\alpha \}].\label{eq:scheme}
\end{multline}\\

\noindent The detailed derivation is in the \emph{Appendix}. {\color{blue}We are
interested in the distribution of the difference between the total energy of the
unperturbed ground state and that of the new ground state at external field $\delta\gamma$,
that is $E_{\beta}-E_{\gs} = U_{\beta}-\frac{\delta\gamma}{\sqrt{N}}\Sigma_{\beta}-U_{\gs}$.
On the other hand, in the calculations it is clear that the relevant variable, whose
distribution can be computed easily, is $\Delta E\equiv U_{\beta}-U_{\gs}-\frac{\delta\gamma}{\sqrt{N}}\left[\Sigma_{\beta}-\Sigma_{\gs}\right] =
E_{\beta}-E_{\gs}+\frac{\delta\gamma}{\sqrt{N}}\Sigma_{\gs}$; this is not a
great issue because we are interested in the small $\delta\gamma$ regime, and $\Sigma_{\gs}$
is also a small quantity. 
}
In the full-\RSB{} limit ($k\to\infty$) the result is

\begin{equation}
\label{eq:final}
\fl \Prob[\Delta E | \delta\gamma] = \delta(\Delta E)\mathcal{R}(0,\delta\gamma) - \theta(-\Delta E)\frac{\dd \mathcal{R}(\Delta E,\delta\gamma)}{\dd\Delta E},
\end{equation}
\begin{equation}
\label{eq:functionr}
\fl \mathcal{R}(\Delta E,\delta\gamma) = \exp\left \{-\abs{\delta\gamma} \int \dd q\ y^\prime(q)\sqrt{1-q} \cdot \rho\left[\frac{\Delta E+\delta\gamma^2 Y(q)}{\sqrt{1-q} \abs{\delta\gamma}}\right] \right \},
\end{equation}

\noindent where $\rho(x)\ \equiv\ \frac{e^{-\frac{x^2}{4}}}{\sqrt{\pi}} + \frac{x}{2}\,\mathrm{erfc}\left(-\frac{x}{2}\right)${\footnote{$\mathrm{erfc}(x) = \frac{2}{\sqrt{\pi}}\int_x^\infty e^{-t^2}\dd t$}} and $Y(q)\equiv \int_q^1\dd q^\prime \, y(q^\prime)$.
Keep in mind that by definition $\Delta E \leq 0$, since the new ground state has
to be lower than the unperturbed one; notice also the manifest invariance under
$\delta\gamma \rightarrow -\delta\gamma$, {\color{blue}that arises naturally from the computations
and suggests the presence of a cusp for $\delta\gamma=0$. In principle, for different
kinds of perturbations this symmetry might not hold.}

\vspace{2em}
\section{Asymptotic behavior}

The increasing function $y(q)$ is related to the distribution of overlaps between states.
In some cases, a large concentration of states close to the ground state causes
the function $y^\prime(q)$ to diverge near $q=1$, as $y^\prime(q)\sim {(1-q)}^{-\mu-1}$:
for instance, {\color{blue}systems of soft spheres display a divergence with exponent $\mu_\mathtt{J} \approx \frac{1}{1.41}$ (in
literature it is called also $\frac{1}{\kappa}$) at jamming, and $\mu_\UNSAT{} = \frac{1}{2}$ in the \UNSAT{} phase,}
and the Sherrington-Kirkpatrick model also has an exponent
$\mu_\mathtt{SK} = \frac12$. The accumulation of neighboring states has
consequences on the distribution of avalanches, since one can imagine that the
system will jump easily even with small perturbations: indeed, we are going to
show that the probability distribution develops a power-law behavior for sufficiently
small jumps and fields, with an exponent that is directly linked to the exponent $\mu$.\\ \\
In order to study the asymptotic behavior, let us introduce the function
$\mathcal{C}(\Delta E, \delta\gamma) \equiv -\log\mathcal{R}(\Delta E,\delta\gamma)$,
where $\mathcal{R}$ is the same function as in \Eref{eq:final}.
{\color{blue}The probability density of jumps is
$\Prob[\Delta E|\delta\gamma] = \mathcal{R}(\Delta E, \delta\gamma)\ \partial_{\Delta E} \mathcal{C}(\Delta E, \delta\gamma)$;
for $\abs{\Delta E} \ll \abs{\delta\gamma} \ll 1$, we have}
\begin{multline}
  \partial_{\Delta E}\mathcal{C}(\Delta E, \delta\gamma) = \frac12 \int_0^1\dd q\, y^\prime(q)\ \mathrm{erfc}\!\left[-\frac{\Delta E + \delta\gamma^2 Y(q)}{2\sqrt{1-q}\abs{\delta\gamma}}\right] \sim \\
\sim \frac12\abs{\frac{\Delta E}{\delta\gamma}}^{-2\mu} \int_{\abs{\frac{\Delta E}{\delta\gamma}}^2}^\infty \dd u\, u^{\mu-1}\mathrm{erfc}\!\left(\frac{\sqrt{u}}{2}\right) \sim \abs{\frac{\Delta E}{\delta\gamma}}^{-2\mu}.
\end{multline}
\noindent Integrating and exponentiating we find also the behavior of
$\mathcal{R}(\Delta E, \delta\gamma)$, and, in the end, the asymptotic behavior of $\Prob[\Delta E, \delta\gamma]$:
\begin{equation}
\label{eq:exponent}
\fl\Prob[\Delta E|\delta\gamma] \sim \left \{\begin{array}{lr}
    \exp\left \{-\mathrm{const}\times \abs{\delta\gamma} \abs{\frac{\Delta E}{\delta\gamma}}^{-2\mu+1}\right \} \abs{\frac{\Delta E}{\delta\gamma}}^{-2\mu}, & \text{for } \mu>\frac12,\\
    \abs{\frac{\Delta E}{\delta\gamma}}^{-1+\mathrm{const}\times \abs{\delta\gamma}}, & \text{for } \mu=\frac12.\\
    \end{array} \right.
\end{equation}

\begin{figure}[h!]
\centering
\includegraphics[width=0.45\textwidth]{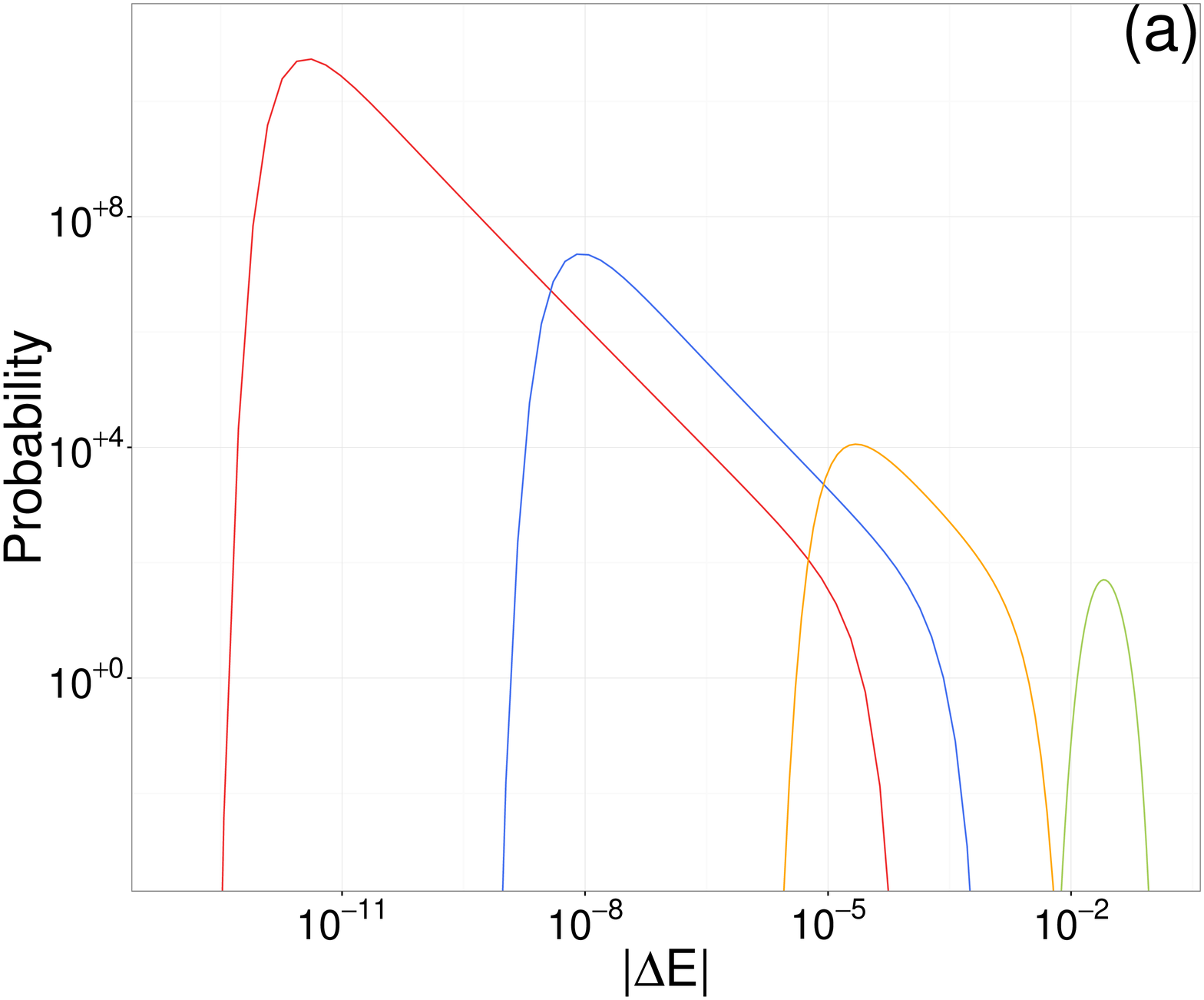}
\includegraphics[width=0.45\textwidth]{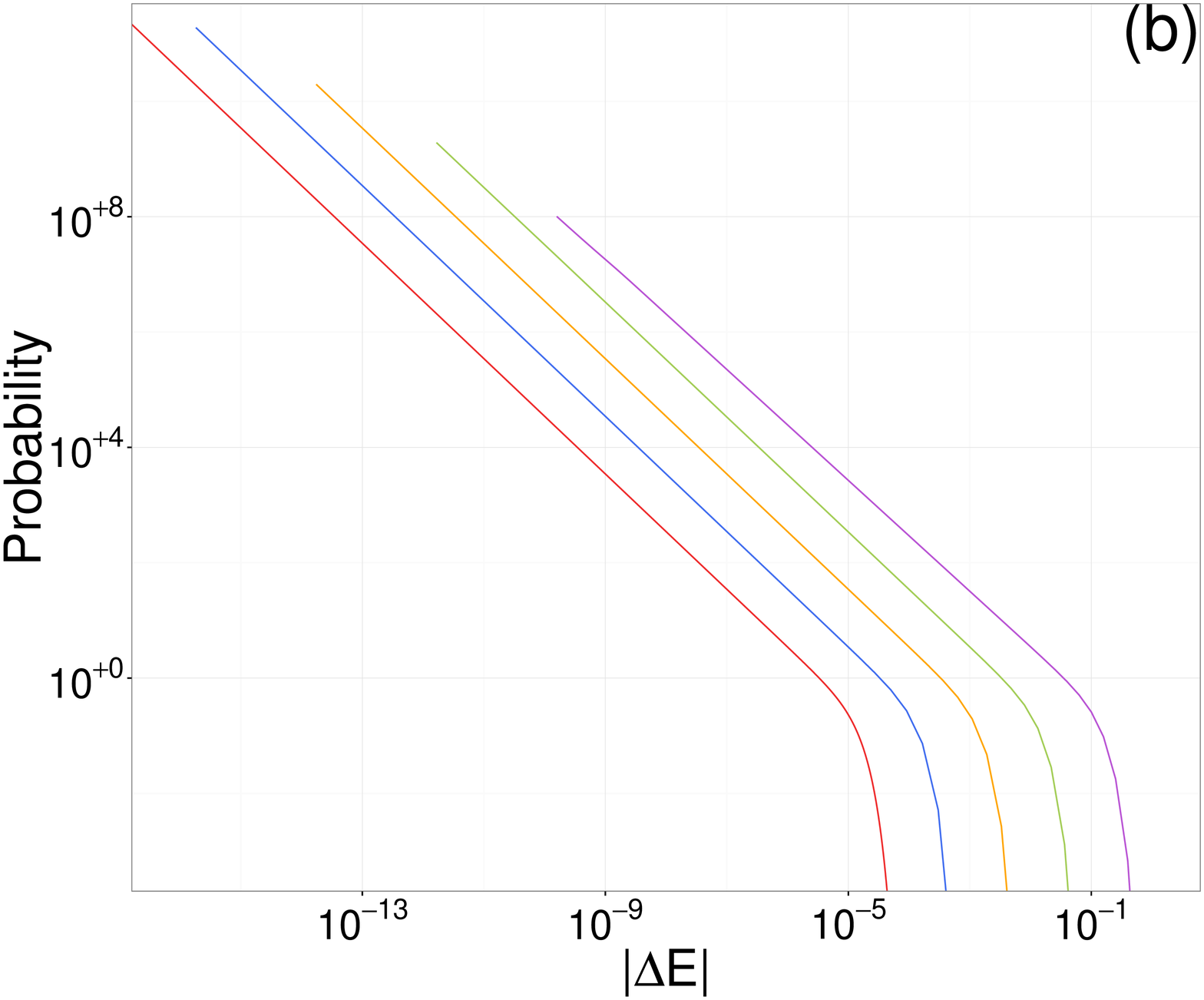}
\caption{{\bf a.} On the left, the plot of the distribution for a tentative function
$y(q)$ that diverges with an exponent $\mu = \mu_\mathtt{J}\approx \frac{1}{1.41}$,
for several values of the perturbing field: notice the development of the power-law
region as the field is lowered. {\bf b.} On the right, the plot for a function $y(q)$
that diverges with $\mu = \mu_\mathtt{SK} = \mu_\mathtt{UNSAT} = \frac12$, for several perturbations;
here, the exponent of the power law displays minor corrections for larger field.\label{fig:analyticplots}}
\end{figure}

\noindent Notice how, for $\mu>\frac12$, $\Prob[\Delta E|\delta\gamma]$ is a power
law if $\abs{\delta\gamma^{1+\frac{1}{2\mu-1}}} \ll \abs{\Delta E} \ll \abs{\delta\gamma} \ll 1$;
therefore, the \emph{avalanche exponent} is found to be $\tau\equiv 2\mu$. {\color{blue}For
$\mu=\frac12$ (and small field $\delta\gamma$) there is a small correction to the exponent, of order $|\delta\gamma|$.} In
\Fref{fig:analyticplots} are shown the plots of the probability distribution related
to two different functions $y(q)$, for some small values of the field.
{\color{blue}We can compare our result with~\cite{ledoussal2012equilibrium}: in that article
the authors find, via a differential equation approach based on replica symmetry
breaking, that the density of static avalanches \emph{in stress} (they actually use
the language of magnetic systems and call it \emph{magnetization}) per unit $\delta\gamma$ is given by
\begin{equation}
\mathcal{P}[\Delta\Sigma] = \theta(\Delta\Sigma) \Delta\Sigma \int_0^1\dd q\,y^\prime(q) \frac{e^{-\frac{\Delta\Sigma^2}{4\pi(1-q)}}}{4\pi(1-q)}.\label{eq:ledoussal}
\end{equation}
If $y(q)$ diverges, then the integral is dominated by $q\approx 1$ and the probability
displays a power-law behavior for small jumps, with an exponent $\tau=2\mu$ like
for the jumps in the total energy. In our framework we can recover the same result
starting from~\Eref{eq:prefinal} in the appendix (or, conceptually, \Eref{eq:scheme});
this equation, valid for a finite $k$-\RSB{}, defines the probability distribution $\mathcal{P}[\Delta U,\Delta\Sigma,q|\delta\gamma]$,
where $\Delta U, \Delta\Sigma$ are the differences in energy and stress between the unperturbed
and perturbed ground states, and $q$ is their mutual overlap. If we keep only the first
order in the Taylor expansion for small fields $\delta\gamma>0$, then integrate out
the energy jump $\Delta U$ and the overlap $q$, and divide by the field strength
$\delta\gamma$, we find the density of stress jumps per unit strain, a result
identical to~\Eref{eq:ledoussal}. Analogously, integrating out $\Delta U$ and $\Delta\Sigma$
and dividing by $\delta\gamma$ we can find the density $\mathcal{P}[q]$ of jumps
at a given overlap $q$:
\begin{equation}
\mathcal{P}[q]=\sqrt{\frac{1-q}{\pi}} y^\prime(q)\sim {(1-q)}^{-\mu-\frac12}.
\end{equation}
In the case of the Sherrington-Kirkpatrick model studied in~\cite{ledoussal2012equilibrium}
we find the same result presented in the article, namely that $\mathcal{P}[q]\sim\frac{1}{1-q}$.
For completeness, with the same approach (expanding for small $\delta\gamma$, integrating out
$\Delta\Sigma$ and $q$, and dividing by $\delta\gamma$), we find the density of jumps
in internal energy per unit field,
\begin{equation}
\mathcal{P}[\Delta U]\sim\frac{1}{\delta\gamma}{\left(\frac{\Delta U}{\delta\gamma}\right)}^{-2\mu}.
\end{equation}
where the fact that $\delta\gamma$ did not disappear is because the energy jumps are of
order $\delta\gamma$.}

\noindent{\color{purple}It is interesting to discuss what happens when the perturbing field $\delta\gamma$ scales
as $N^{-\alpha}$ for some exponent $\alpha \geq 0$. If $\alpha=0$ (i.e. $\frac{\delta\gamma}{\sqrt{N}} \sim N^{-\frac12}$) then $\mathcal{R}(0,\delta\gamma) = 0$,
where $\mathcal{R}$ is the function in~\Eref{eq:functionr};
since $\mathcal{R}(0,\delta\gamma)$ is the probability of not jumping when a shear strain
$\delta\gamma$ is applied, this scaling implies that, in the thermodynamic limit,
the system jumps with any perturbation, however small it might be. Moreover,
the power-law behavior in the distribution of the jumps $\Delta E$ is suppressed
beyond the region $|\delta\gamma|^{1+\frac{1}{2\mu-1}} \ll |\Delta E| \ll |\delta\gamma|$,
and the typical jump is of order $|\Delta E| \sim N^0$ (remember that this is the
sub-leading, non-extensive part of the free-energy, and that the extensive term is the same for all the states).\\ \\
There is, possibly, another interesting regime, that is the one that leads to a finite probability
of not jumping when a small shear strain is applied, even when $N\to\infty$. The zero-temperature maximum overlap between two
states --- the Edwards-Anderson order parameter $q_\mathrm{EA}$, --- is 1 in the thermodynamic limit,
and it might, in principle, scale as $1 - N^{-\beta}$. In this case, all the diverging integrals in $\dd q$ have
a cut-off; scaling $\delta\gamma$ as $N^{-\alpha}$ with $\alpha = \frac12\beta(2\mu-1)$ leads to
a finite probability for the system not to jump when sheared, but now the typical jump
is smaller and scales as $|\Delta E| \sim N^{-\beta(2\mu-1)}$.}

\vspace{2em}
\section{Simulations of systems of spheres}

We would like now to compare the prediction of replica theory with numerical
simulations in three dimensions. As stated in the introduction, at zero temperature
one finds different critical scalings at jamming (packing fraction $\phi\equiv\phi_J$) and
in the \UNSAT{} regime $\phi>\phi_J$.
As described in~\cite{spheresIII} the jamming solution is characterized by a singular
$y_J(q)\sim {(1-q)}^{-\mu_J}$ for $q\to 1$. The exponent $\mu_J$ is related to the ``pseudo-gap'' exponent $\theta$ in the
distribution of small contact forces at jamming, $P(F)\sim F^\theta$, according to
\begin{eqnarray}
  \label{eq:1}
  \mu_J=\frac{3+\theta}{2(2+\theta)},
\end{eqnarray}
where the numerical value predicted for $\theta$ is $\theta\approx 0.42311$. Accordingly,
denoting $\tau_J$ the value of the avalanche exponent $\tau$ at jamming, we have $\tau_J=\frac{3+\theta}{2+\theta}\approx 1.41269$.
On the other hand a different solution has been found in the \UNSAT{} phase above
jamming~\cite{FranzParisiSevelevUrbaniZamponi} which predicts $y(q)\sim {(1-q)}^{-1/2}$
for $q\to 1$ and correspondingly denoting $\tau_\UNSAT{}$ the avalanche exponent in this
region, we have $\tau_\UNSAT{}=1$. Notice that this same scaling is observed in the
Sherrington-Kirkpatrick spin glass model ($\mu_\texttt{SK}=\frac12$).\\ \\
We then study the distribution of avalanches in numerical simulations of systems of 3-dimensional
soft spheres under shear strain. We consider the standard mono disperse harmonic
soft sphere model with a potential between spheres at distance $r$
\begin{eqnarray}
  \label{eq:2}
V(r)={\left(1-\frac{r}{2R}\right)}^2 \theta\left(1-\frac{r}{2R}\right)
\end{eqnarray}
where $R$ is the radius of the particles and $\theta(x)$ is the step function.
We first prepare the system generating amorphous energy minima at packing fractions
$\phi\geq \phi_J$, and then we study the effect of an applied shear according to
the athermal quasi-static protocol, \textcolor{blue}{letting the system relax after
every perturbation via a dissipative molecular dynamics~\cite{bitzek2006structural}}.
The jammed systems are prepared using compression/decompression cycles that end
when the average number of contacts per particle reaches the desired isostatic value (twice the dimension),
and lead to a jamming packing fraction $\phi_J\approx 0.64$; {\color{blue}\UNSAT{} samples are
prepared directly at some target packing fraction $\phi>\phi_J$.} \textcolor{blue}{The shear-strain is achieved
using the \emph{Lees-Edwards boundary conditions}~\cite{leesedwards};
these boundary conditions shift each periodic unit by an amount
that is proportional to its vertical position with respect to the central unit.}
{\color{blue}In principle our theoretical work computes the statistics of the jump
starting from the ground state, induced by a small external field $\delta\gamma$;
in practice we increase the strain by small steps $\delta\gamma$ up to a small maximum accumulated strain
$\gamma_{\max}=0.01$, and the energy is minimized between every step, while recording
all the observables and computing the jumps between to subsequent steps.
This is justified by the assumption of stationarity in the linear
stress-strain regime: for this reason the value of $\gamma_{\max}$ is kept small in order not to
perturb too much the system.}

\begin{figure}[h!]
\centering
\includegraphics[width=0.9\textwidth]{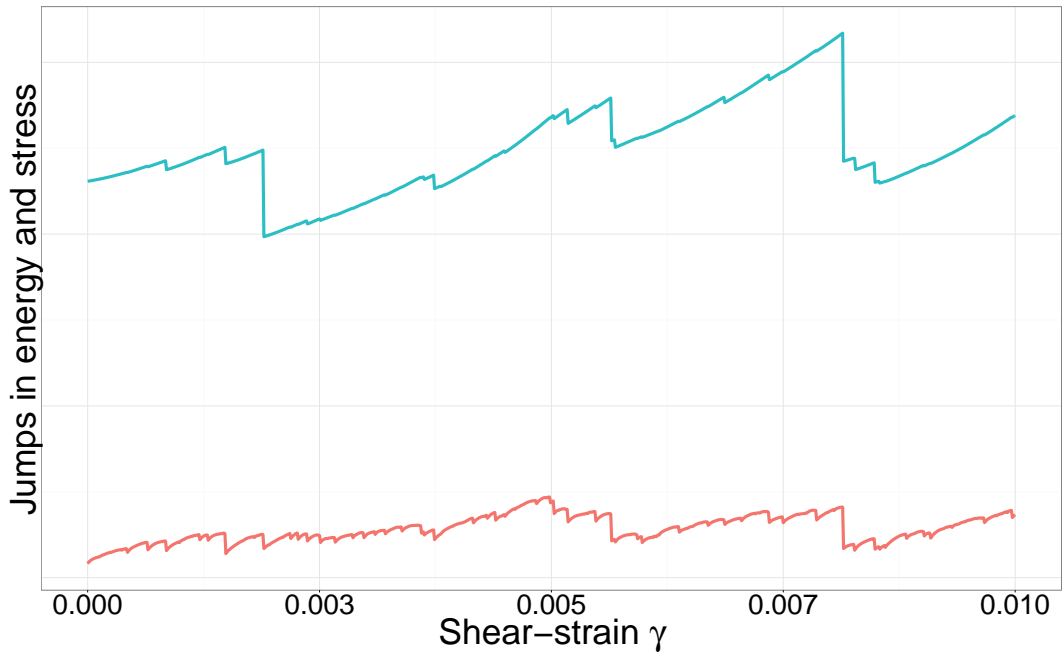}
\caption{Energy and stress as a function of the accumulated shear strain for a system
of $N=4000$ particles at jamming ($\phi\approx0.64$). Every step is made with $\delta\gamma=10^{-5}$.
We are interested in the distribution of jumps $E_{\min}(\gamma+\delta\gamma)-E_{\min}(\gamma)$,
{\color{blue}$E_{\min}(\gamma)$ being the instantaneous ground state when subject to a perturbation $\gamma$}.\label{fig:jumps}}
\end{figure}

\noindent We detect the avalanches in the systems by measuring the energy as
a function of strain as in \Fref{fig:jumps}. We measure the avalanche distribution
at various values of the packing fraction, at jamming and in the \UNSAT{} region
($\phi=0.64,0.75,0.8,0.9$) and for different values of the system size ($N=500,1000,2000,4000$
particles), in three dimensions. We generate several hundreds of configurations
(about 300 at jamming and 1000 at higher packing fractions) for each value of $N$
and $\phi$; then, every sample is sheared for 1000 steps with strain increment
$\delta\gamma=10^{-5}$ and we record the amplitude of the energy jumps
$E(\gamma+\delta\gamma) - E(\gamma) - \delta\gamma\Sigma(\gamma)$ ($\approx E(\gamma+\delta\gamma) - E(\gamma)$
since both $\delta\gamma$ and $\Sigma(\gamma)$ are very small in the regime that we are simulating).
\begin{figure}[h!]
\centering
\includegraphics[width=0.45\textwidth]{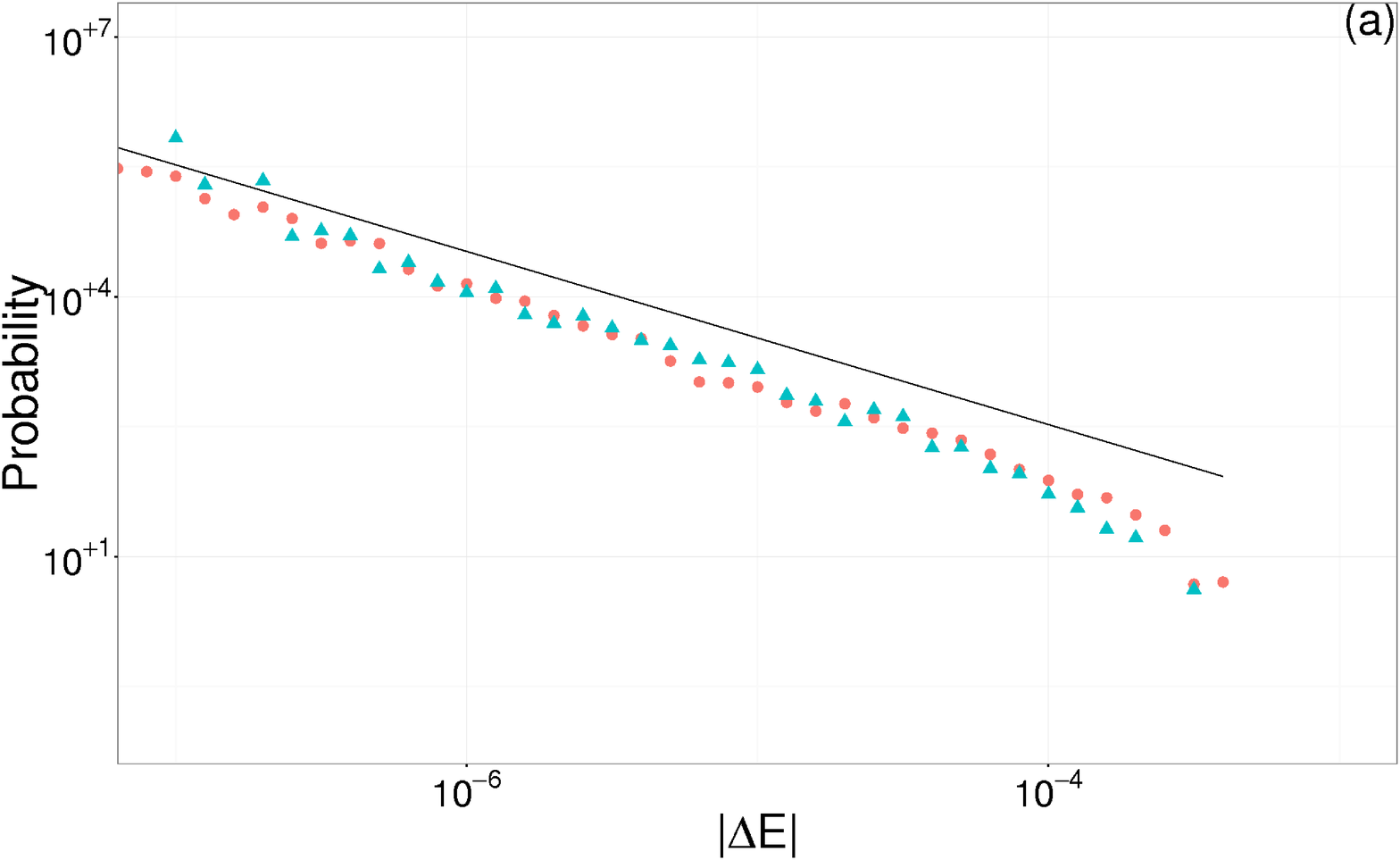}
\includegraphics[width=0.45\textwidth]{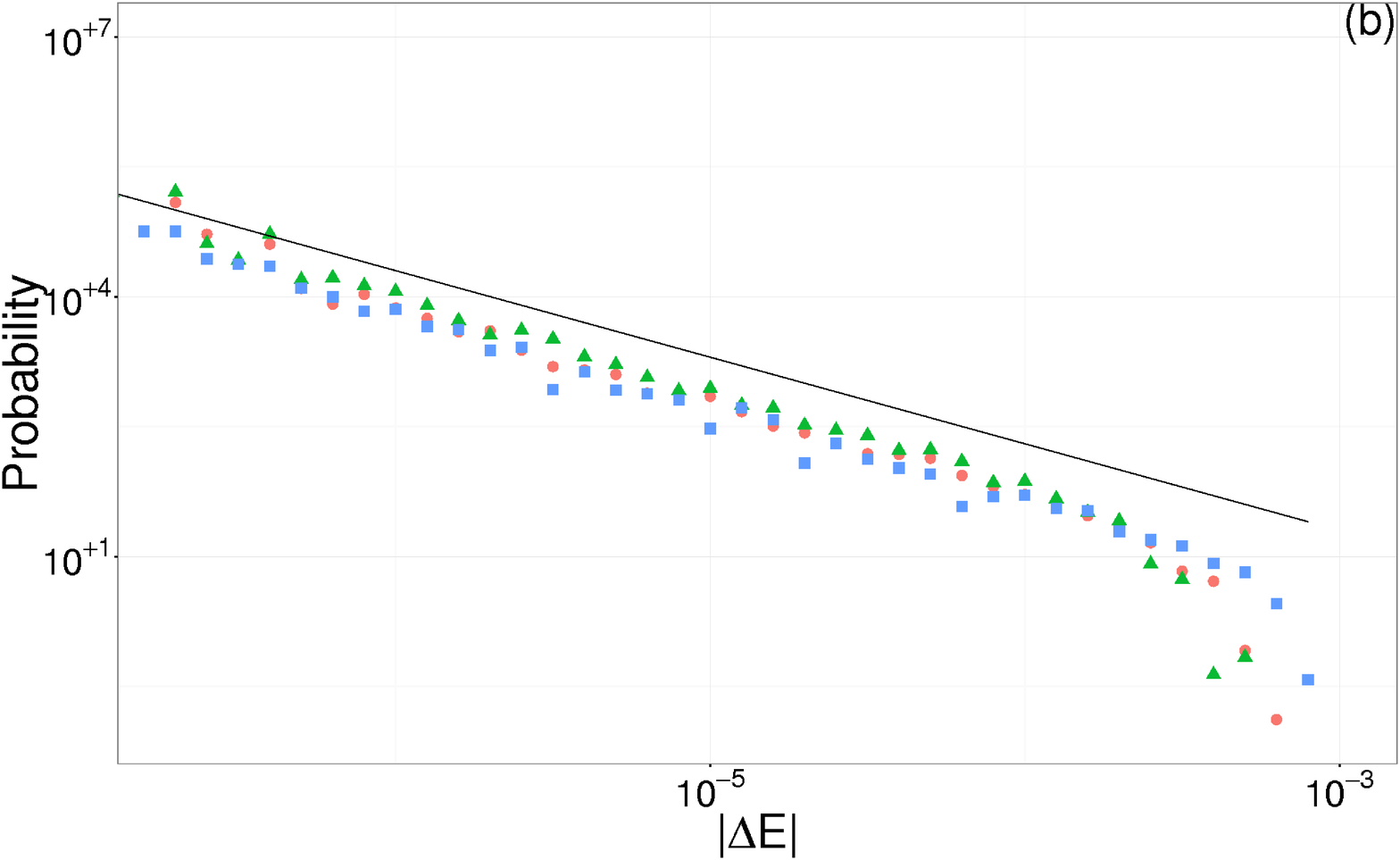}\\
\includegraphics[width=0.45\textwidth]{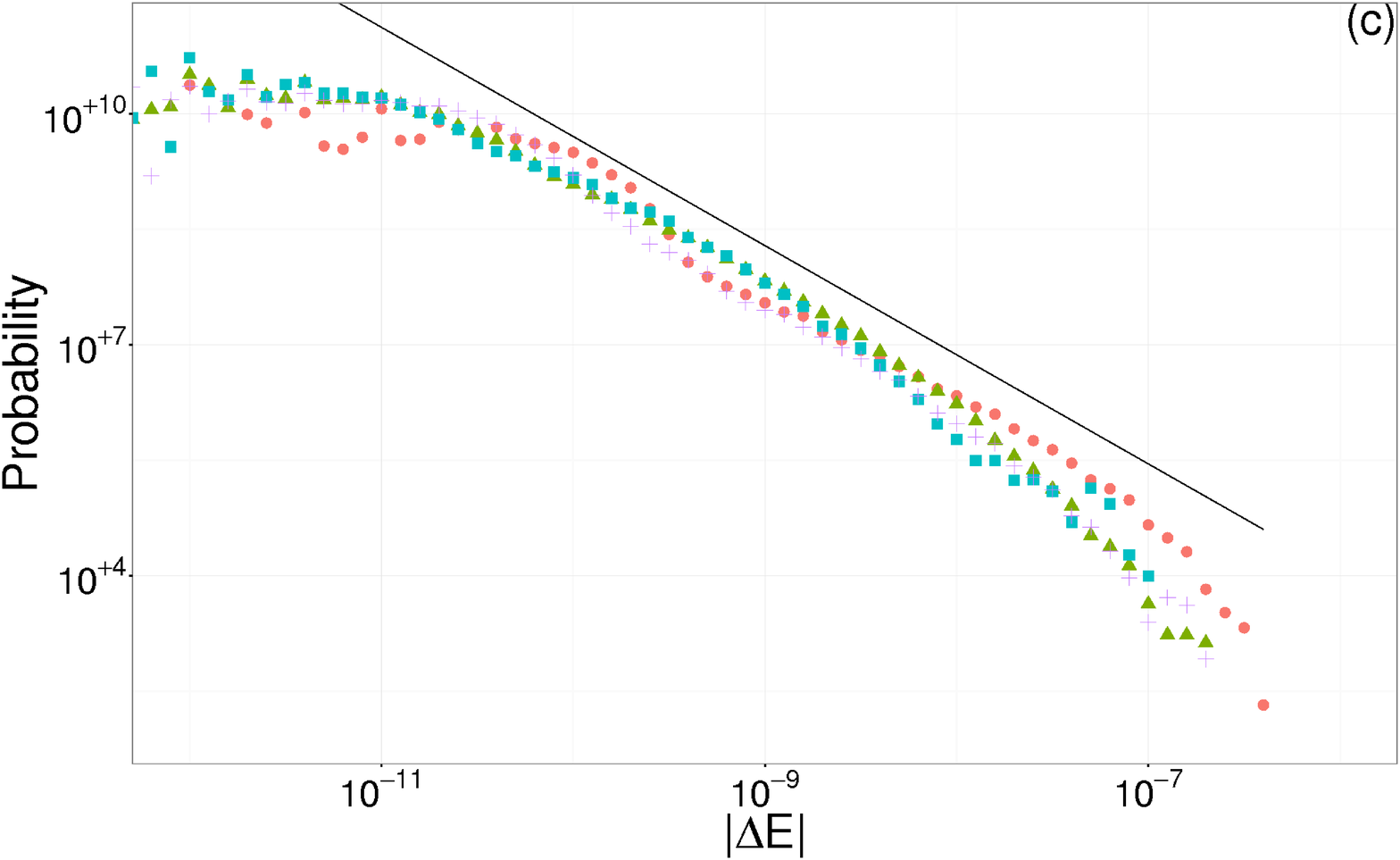}
\includegraphics[width=0.45\textwidth]{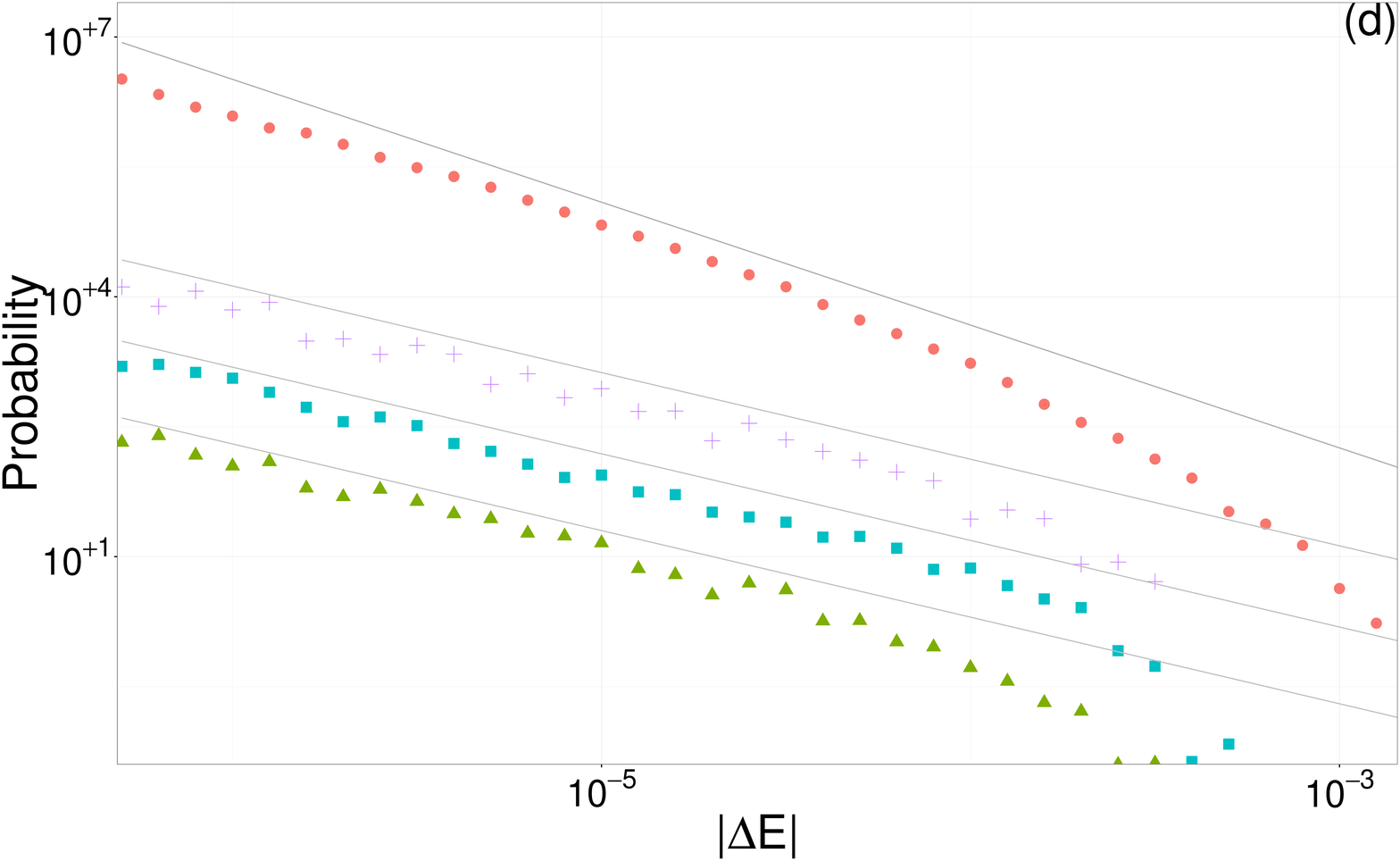}
\caption{Histograms for the avalanche distribution in systems of several sizes,
with $\delta\gamma = 10^{-5}$ (shifted in order to be compared with the predicted power-law).
Histograms of the avalanches for systems
in the \UNSAT{} phase, at prepared at packing fractions \textbf{(a)} $\phi=0.75$
(the fitted exponent is $\tau_{0.75,\mathrm{fit}} \approx 1.11\pm 0.029$)
and \textbf{(b)} $\phi=0.80$ (the fitted exponent is $\tau_{0.80,\mathrm{fit}} \approx 1.12\pm 0.028$).
\textbf{(c)} Jammed system at $\phi\approx 0.64$ (the fitted exponent is $\tau_{J,\mathrm{fit}} \approx 1.52\pm 0.075$).
\textbf{(d)} Comparison of all the data.
The predicted power laws with $\tau_J\approx\frac{2}{1.41}\approx 1.42$ and
$\tau_\UNSAT{}=1$ are also shown in.\label{fig:numericalplots}}
\end{figure}
In \Fref{fig:numericalplots} we present the histogram of the energy jumps in a
log-log plot, showing that a power law regime exists both at jamming and in the
jammed configurations. It is manifest that the exponent in the jammed phase is
smaller than the one at jamming, and it is independent of the value of $\phi$.
\textcolor{blue}{The deviation from the theoretical behavior presented in \Fref{fig:analyticplots}
has several origins: first of all, the theoretical curves are strictly valid for infinite
dimensional systems, and even though we can argue that some properties do not vary with the
dimension, there are features that surely do; then, the scale of the shear-strain
axis depends on a constant factor in the variance of the shear-stress
in the states; furthermore, there is a Gaussian error in the computation of the
ground states due to the minimization algorithm, that ends when some precision
criteria are met: this error is apparent only at small scales, that can be seen
in the low-strain part of the histogram at jamming. Among these sources of errors,
only the first one (the fact that we are comparing finite and infinite dimensions)
might have an effect on the exponent, but surprisingly it does not --- at least not by much.}
The quality of our data does not allow a proper determination of the
avalanche exponent, however in both cases the data are compatible with the
theoretical values; we have presented our numerical fits in the caption of~\Fref{fig:numericalplots}.
While this was expected at jamming it is a surprising
result in the jammed phase in finite dimensions. More accurate simulations
in different dimensions should be performed to validate this result.

\vspace{2em}
\section{Consequences on the elastic moduli}

{\color{purple}Systems compressed at or above jamming attain some mechanical stability
due to the contacts between particles (e.g.~\Fref{fig:phiz}). A question that arises
naturally is whether the elastic response of such systems shares common properties
with that of ordered solids. Crystals, for instance, when sheared, display at first
an elastic (linear) response in the shear stress, that at a certain point (the \emph{yielding point}~\cite{lin2014scaling})
saturates; this happens because for large shear strains the system \emph{fails} and
displays a \emph{plastic}, irreversible response. Interestingly, disordered systems,
\emph{on average}, behave in a very similar manner: as one can see from~\Fref{fig:yielding},
the average stress is a linear function of the shear-strain for small perturbations (for this system
of soft spheres, $\gamma \lesssim 0.02$), but it becomes a constant for larger values.\\ \\
Avalanche dynamics provides the description of the response of the system at the microscopic
level, and in our picture its statistics is a consequence of the glassy criticality
associated with the divergence of the Parisi function $y(q)$ near $q=1$. It has been
recently pointed out in~\cite{biroli2016breakdown} that \emph{at the transition between
stable and marginally stable glasses} (namely the \emph{Gardner} transition~\cite{gardner1985spin})
the elastic response exhibit a singular behavior.

\begin{figure}[hb]
  \includegraphics[width=\textwidth]{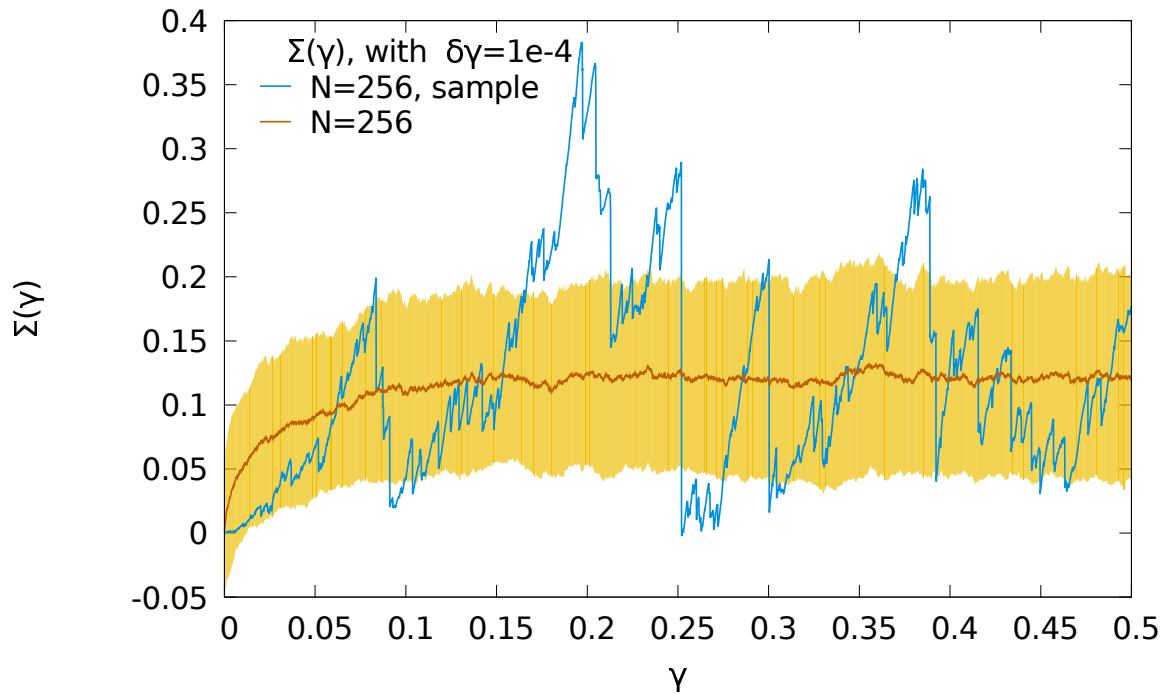}
  \caption{Average stress-strain curve for a system of 256 soft spheres (average over
  200 samples); the shaded area shows the associated sample-to-sample variance.
  Superimposed is also shown the stress-strain curve of typical sample.}\label{fig:yielding}
\end{figure}

\noindent In this section we would like to discuss the relation between the macroscopic
elastic response and the distribution of avalanches, and show that singular responses
naturally emerge in zero-temperature glassy phases and at the jamming point.
Let's say that a given sample, when sheared, produces a stress curve $\Sigma(\gamma)$;
for small $\gamma$ (that is, in the elastic, linear regime) the distribution of $\Sigma(\gamma)$ (relative to the initial value of the stress)
is given by the distribution $\Prob[\Sigma(\gamma)]\equiv\Prob[\Sigma|\gamma]$ of stress jumps that we have found.
Ideally, one would like to expand $\Sigma(\gamma)$ in powers of $\gamma$, in such a way
to define higher (non-linear) elastic moduli, and then compute their first moments.
We can define similar quantities by considering the moments of the \emph{finite
differences} of this curve, and then taking the suitable limit:

\begin{equation}
  \left<\mu_m^k|\gamma\right> \equiv \lim_{\delta\gamma\rightarrow 0} {\delta\gamma}^{-km} \left<{\left[\sum_{n=0}^m {m \choose n} {(-1)}^{(m-n)}\,\Sigma(\gamma+n\delta\gamma)\right]}^k\right>.
  \label{eq:generalmoduli}
\end{equation}

\noindent For example, the first and second moments of the first (shear) modulus are

\begin{equation}
  \left<\mu_1|\gamma\right> \equiv \lim_{\delta\gamma\rightarrow 0} {\delta\gamma}^{-1} \left<\Sigma(\gamma+\delta\gamma)-\Sigma(\gamma)\right> = \partial_\gamma \left<\Sigma|\gamma\right>,
  \label{eq:shearmod1}
\end{equation}
\begin{equation}
  \left<\mu_1^2|\gamma\right> \equiv \lim_{\delta\gamma\rightarrow 0} {\delta\gamma}^{-2} \left<{\left[\Sigma(\gamma+\delta\gamma)-\Sigma(\gamma)\right]}^2\right>.
  \label{eq:shearmod2}
\end{equation}

\noindent Of course for $k=1$ we recover $\left<\mu_m|\gamma\right> = \partial_\gamma^m\left<\Sigma(\gamma)\right>$,
that in the linear regime is always finite. In order to compute $\left<\mu_1^2|\gamma\right>$,
we have to make an assumption of \emph{stationarity}; what we will assume is that
the distribution of $\Delta\Sigma \equiv \Sigma(\gamma+\delta\gamma)-\Sigma(\gamma)$
does not depend on $\gamma$ (this assumption seems reasonable in the elastic regime,
when $\gamma\ll 1$). Therefore $\left<\mu_1^2|\gamma\right> = \lim{}\, {\delta\gamma}^{-2} \left<\Delta\Sigma^2|\delta\gamma\right> = \infty$,
since we can show that $\left<\Delta\Sigma^2|\delta\gamma\right> \sim \delta\gamma$
--- for example starting from~\Eref{eq:prefinal}, --- either at jamming or in the \UNSAT{} phase.\\

\begin{figure}[hb]
  \includegraphics[width=0.5\textwidth]{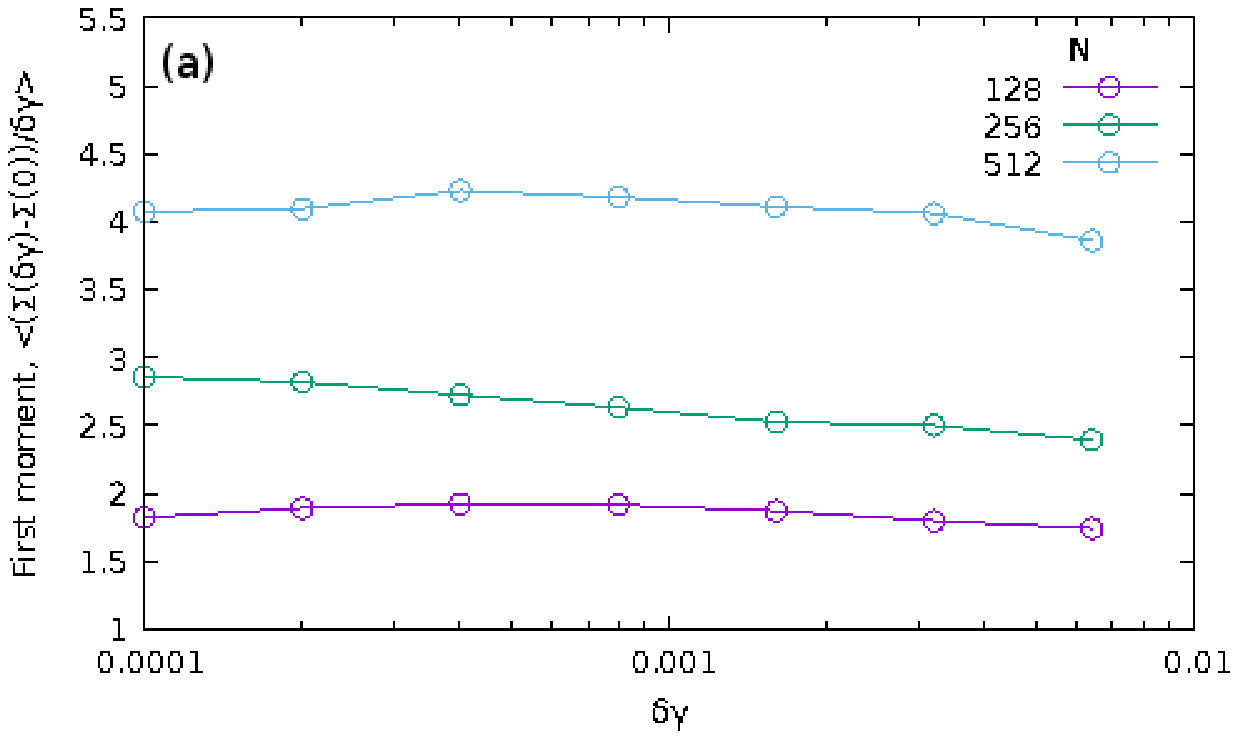} 
  \includegraphics[width=0.5\textwidth]{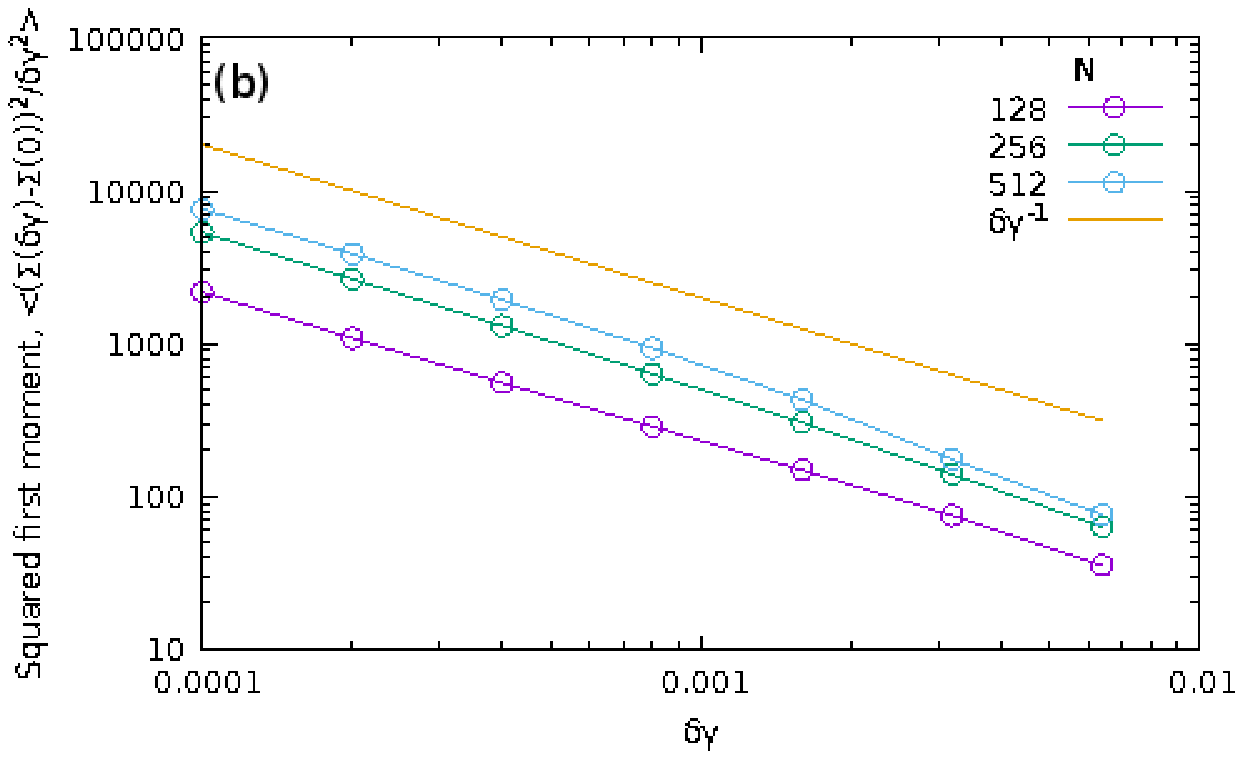}\\ 
  \includegraphics[width=0.5\textwidth]{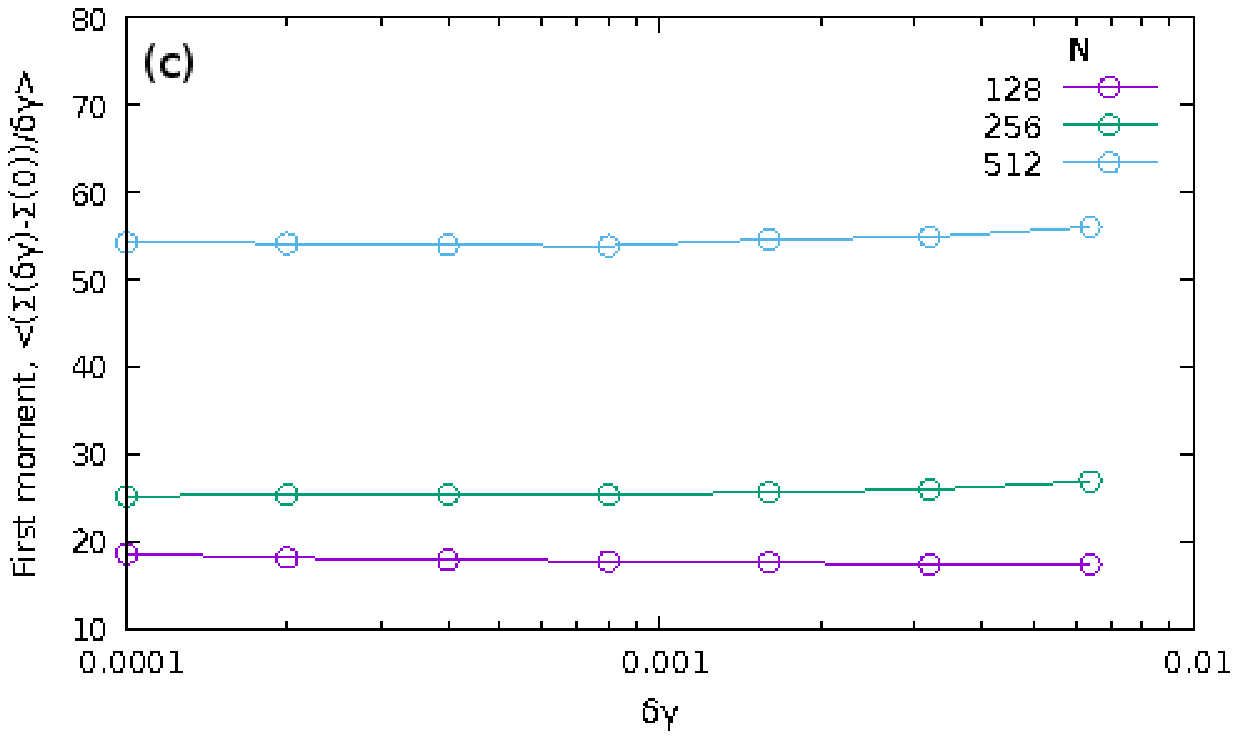} 
  \includegraphics[width=0.5\textwidth]{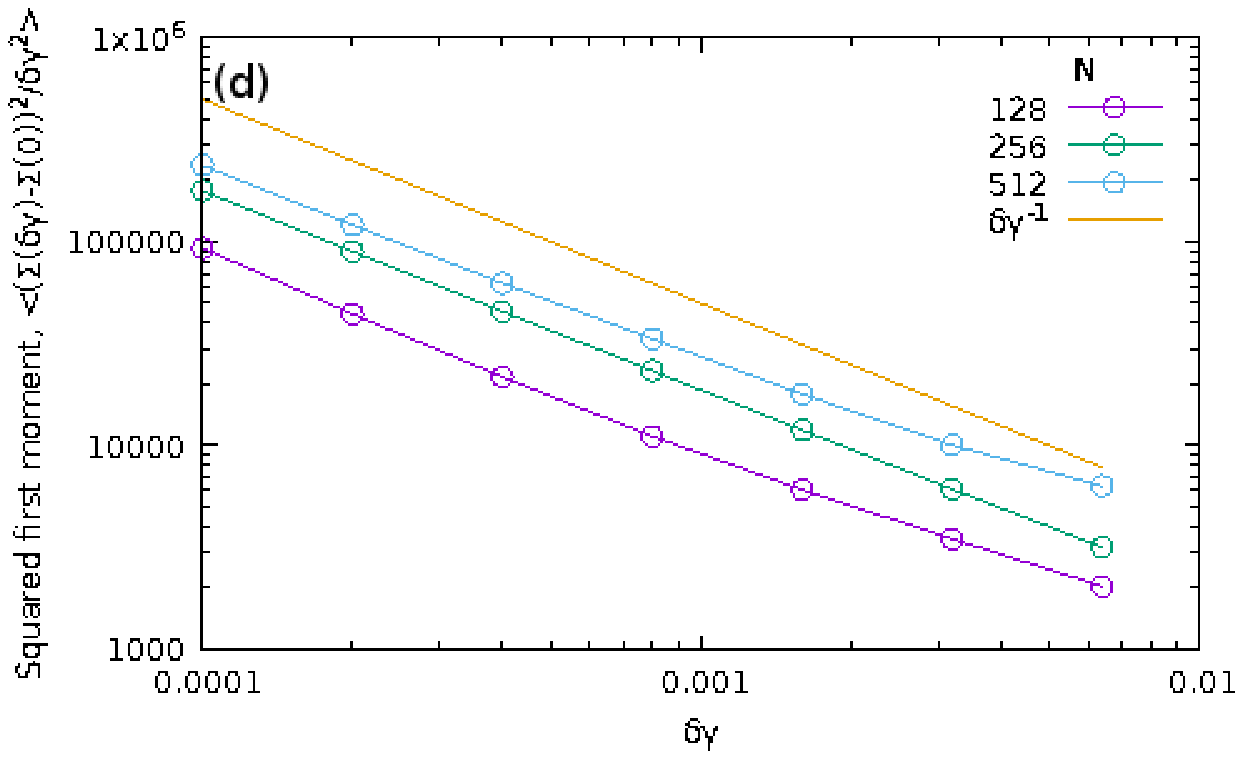}\\ 
  \caption{Behavior of the first two moments of the shear modulus $\mu_1$ as a function of $\delta\gamma$.
  \textbf{(a)} and \textbf{(c)} show the average shear modulus (at jamming and at $\phi=0.75$,
  respectively), that is finite in the $\delta\gamma\to 0$ limit. \textbf{(b)} and \textbf{(d)}
  show the sample-to-sample variance of the shear modulus (again at jamming and at $\phi=0.75$);
  both diverge in the $\delta\gamma\to 0$ limit as $\delta\gamma^{-1}$, as predicted.
  Log scale on the $\delta\gamma$ axis in both figures; log scale on the vertical axis only in the pictures to the right.}\label{fig:numericalmoments}
\end{figure}

\noindent Our numerical simulations confirm this result: in~\Fref{fig:numericalmoments} are shown
$\left<\mu_1|\gamma\right>$ and $\left<\mu_1^2|\gamma\right>$ before taking the limit $\delta\gamma\rightarrow 0$
in~\Eref{eq:shearmod1}: the two quantities are computed with a finite $\delta\gamma$, in order to
extrapolate the proper limit; $\left<\mu_1^2|\gamma\right>\sim\delta\gamma^{-1}$ in both phases, as predicted.
In~\cite{biroli2016breakdown} the authors have found that systems at the Gardner
transition have a shear-modulus with finite variance ($\left<\mu_1^2|\gamma\right> < \infty$),
but the variance of all the higher moduli diverge. In this work we have shown that
also beyond the Gardner transition (namely, at jamming or in the \UNSAT{} phase) the
elastic response breaks down; even more drastically, in these phases the variance
of the shear modulus is divergent, too.

}

\vspace{2em}
\section{Summary and conclusions}

{\color{blue} In this paper we reexamined the problem of avalanches in mean-field
glassy systems with particular attention to marginal glassy phases (the \UNSAT{} phase) and jamming points.
We focus on static avalanches, namely the difference of the energy or some other observable
between the perturbed and unperturbed ground states. Power-law avalanche distributions
are associated with the proliferation of low-energy relative minima close to the ground state. In the
case of elastic spheres, we find different universality classes for the zero-temperature
marginal glass and the jamming point, where the response to sufficiently small shear strains is
described by non-trivial distributions with different power-law behaviors.\\ \\
These results are compared with the response of \emph{quasi-static} avalanches in three-dimensional
soft spheres, that we studied numerically. Despite obvious differences between the
dynamical protocol and the static calculation, our numerical simulations \emph{(a)}
confirm that the avalanche exponents are different at jamming and in the denser phase, and
\emph{(b)} the values of the exponents are close to the mean-field ones. At jamming
this suggests that the super-universality of the force and gap exponents, i.e.\ their
independence on the spatial dimension and preparation protocol, extends to the dynamical
avalanche exponents. More surprising is the coincidence of the exponents in the glassy
phase where renormalization could be expected even at zero temperature. In order to
clarify this point more precise measures of the exponent $\tau$ and a study as a function
of the space dimension are needed.
Finally we have related the avalanche distribution to the distribution of non-linear
elastic moduli, showing that a singular behavior is present in marginal zero-temperature
glasses and that, since higher moments are divergent, the typical response of the
system is not the average one.
}\\

\noindent Our next projects will follow mainly two directions: firstly, we want to run improved
simulations, aimed to a finer characterization of the avalanches in finite dimensional
systems. Secondly, in this article the maximum shear-strain was kept small in order to
remain in the linear regime, but there is also much interest on the stationary \emph{elasto-plastic regime}
that follows the yielding transition; we plan to study the distribution of avalanches
in this setting, in order to be able to compare our results with all the works that
have been published in the field, for example~\cite{leishangthem2016yielding,lin2015criticality,maloney2004subextensive,dubey2016elasticity,jaiswal2016mechanical,regev2015reversibility,fiocco2013oscillatory,puosi2016plastic,arevalo2014size}.

\vspace{2em}
\ack{}

We thank Pierfrancesco Urbani and Giuseppe Foffi for many useful discussions, and Francesco Zamponi for the deep insight he has kindly provided.
We also thank the anonymous referees for their useful advice.
This work was supported by a grant from the Simons Foundation 454941 to Silvio Franz.\\

\vspace{2em}
\section*{Appendix: calculations}

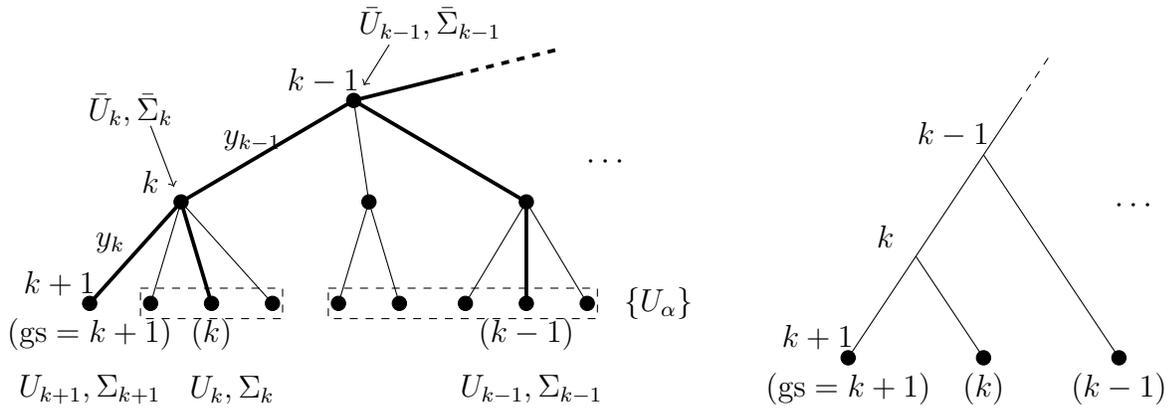
\begin{figure}[hb]
\centering
\begin{tikzpicture}[scale=1.35]

\draw [line width=0.5mm] (0,0) -- (1,0.25);
\draw [line width=0.5mm,dashed] (1,0.25) -- (2,0.5);
\node at (2.5,-0.6) {$\cdots$};

  \draw [line width=0.5mm] (0,0) -- (-1.7,-1);
  \draw [gray] (0,0) -- (0.15,-1);
  \draw [line width=0.5mm] (0,0) -- (1.7,-1);

  \draw [line width=0.5mm] (-1.7,-1) -- (-2.6,-2);
  \draw [gray] (-1.7,-1) -- (-2,-2);
  \draw [line width=0.5mm] (-1.7,-1) -- (-1.4,-2);
  \draw [gray] (-1.7,-1) -- (-0.8,-2);

  \draw [gray] (0.15,-1) -- (-0.15,-2);
  \draw [gray] (0.15,-1) -- (0.45,-2);

  \draw [gray] (1.7,-1) -- (2.3,-2);
  \draw [line width=0.5mm] (1.7,-1) -- (1.7,-2);
  \draw [gray] (1.7,-1) -- (1.1,-2);

  \filldraw (0,0) circle(2pt);
  \filldraw (-1.7,-1) circle(2pt);
  \filldraw (0.15,-1) circle(2pt);
  \filldraw (1.7,-1) circle(2pt);
  \filldraw (-2.6,-2) circle(2pt);
  \filldraw (-2,-2) circle(2pt);
  \filldraw (-1.4,-2) circle(2pt);
  \filldraw (-0.8,-2) circle(2pt);
  \filldraw (-0.15,-2) circle(2pt);
  \filldraw (0.45,-2) circle(2pt);
  \filldraw (2.3,-2) circle(2pt);
  \filldraw (1.7,-2) circle(2pt);
  \filldraw (1.1,-2) circle(2pt);

  \node at (3,-2) {$\{U_\alpha\}$};

  \node at (-0.3,0.2) {$k-1$};
  \node at (-2,-0.8) {$k$};
  \node at (-2.9,-1.8) {$k+1$};

  \node at (0.75,0.75) {$\bar{U}_{k-1},\bar{\Sigma}_{k-1}$};
  \draw [->] (0.35,0.55) -- (0.1,0.15);

  \node at (-2.2,-0.1) {$\bar{U}_k,\bar{\Sigma}_k$};
  \draw [->] (-2,-0.25) -- (-1.77,-0.85);

  \node at (-1,-0.4) {$y_{k-1}$};
  \node at (-2.4,-1.4) {$y_k$};

  \node at (-2.6,-2.3) {$(\mathrm{gs}=k+1)$};
  \node at (-1.4,-2.3) {$(k)$};
  \node at (1.7,-2.3) {$(k-1)$};

  \node at (-2.6,-2.85) {$U_{k+1},\Sigma_{k+1}$};
  \node at (-1.2,-2.85) {$U_k,\Sigma_k$};
  \node at (1.75,-2.85) {$U_{k-1},\Sigma_{k-1}$};

  \draw [dashed] (-2.1,-1.85) -- (-0.7,-1.85) -- (-0.7,-2.15) -- (-2.1,-2.15) -- (-2.1,-1.85);
  \draw [dashed] (-0.25,-1.85) -- (2.4,-1.85) -- (2.4,-2.15) -- (-0.25,-2.15) -- (-0.25,-1.85);
\end{tikzpicture}
\hfill
\begin{tikzpicture}[scale=1.35]
\draw (0,0) -- (1/3,0.5); \draw[dashed] (1/3,0.5) -- (2/3,1);
  \draw (0,0) -- (-4/3,-2); \filldraw (-4/3,-2) circle(2pt);
  \draw (0,0) -- (4/3,-2); \filldraw (4/3,-2) circle(2pt);
  \draw (-2/3,-1) -- (0,-2); \filldraw (0,-2) circle(2pt);

\node at (1.5,-0.5) {$\cdots$};
  \node at (-0.3,0.2) {$k-1$};
  \node at (-2/3-0.3,-0.8) {$k$};
  \node at (-4/3-0.3,-1.8) {$k+1$};

  \node at (-4/3,-2.3) {$(\mathrm{gs}=k+1)$};
  \node at (0,-2.3) {$(k)$};
  \node at (4/3,-2.3) {$(k-1)$};
\end{tikzpicture}
\caption{{\bf Left.} A portion of an ultrametric tree for a $k$-\RSB{} system. The dashed boxes enclose the $q_{k-1}$- and $q_k$-clusters, ($k-1$) and ($k$) are their minima, respectively, and (gs)=$(k+1)$ is the ground state. {\bf Right.} The tree after the integration of the states that are not minima of any cluster, leaving only the relevant branches (thick ones in \Fref{fig:ultrametrictree} \emph{left}); this tree can be thought of as the factor graph of the joint probability of the minima in each cluster $\Prob[U_\gs, \Sigma_\gs; U_k, \Sigma_k; U_{k-1}, \Sigma_{k-1};\dots | \delta\gamma]$.}
\label{fig:appendixultrametrictree}
\end{figure}

\noindent In this section we show how the results were found analytically, via an
approach based entirely on elementary probabilistic methods. {\color{blue}Every state $\alpha$ of the system
has intrinsic stress $\tilde\Sigma_\alpha$ and energy $U_0 + U_\alpha$, where $U_0$
in an extensive term that is identical for all the states in the sample and $U_\alpha$
is a term of order $\mathcal{O}(1)$; since we are interested in finding the state that
minimizes the total energy, from now on we will ignore the extensive term $U_0$.
Anticipating that the distribution of the stresses $\tilde\Sigma_\alpha$ is expected to
be Gaussian in the states, with a variance proportional to the number of particles $N$,
we introduce a set of rescaled stresses $\Sigma_\alpha \equiv \tilde\Sigma_\alpha N^{-\frac12}$
in order to work with $\mathcal{O}(1)$ quantities. Since in the following calculations
$\tilde\Sigma_\alpha$ will always appear multiplied by $\frac{\delta\gamma}{\sqrt{N}}$,
we can get rid of the $N$ terms by considering the total energy of a state as $U_\alpha - \frac{\delta\gamma}{\sqrt{N}} \tilde\Sigma_\alpha \equiv U_\alpha - \delta\gamma \Sigma_\alpha$.}
We assume that the energies $\{U_\alpha \}$ and the stresses $\{\Sigma_\alpha \}$
of a system are distributed according to some probability distribution,
$\Prob_\mathrm{states}[\{U_\alpha,\Sigma_\alpha \}]$; then we want to recover the
marginal distribution of the state that minimizes $U_\beta-\delta\gamma\Sigma_\beta$
among all states at a given perturbation $\delta\gamma$, namely

\begin{multline}
\Prob_{\min}[U_\beta,\Sigma_\beta, \beta | \delta\gamma] =\\
= \int\left[\prod_{\alpha\neq\beta}\dd U_\alpha\dd \Sigma_\alpha \, \theta(U_\alpha - \delta\gamma\Sigma_\alpha > U_\beta - \delta\gamma\Sigma_\beta)\right] \Prob_\mathrm{states}[\{U_\alpha,\Sigma_\alpha \}].
\label{eq:marginal}
\end{multline}\\

\noindent For the sake of clarity, in this section we use the notation $\theta(A>B) \equiv \theta(A-B)$
and $\theta(A<B) \equiv \theta(B-A)$, $\theta(\cdot)$ being the step function.\\ \\
The distribution $\Prob_\mathrm{states}$ is related to the \emph{replica symmetry breaking}
(\RSB) solution; we will be dealing with solutions that are continuously broken (what
is called full-\RSB), but they can be approximated as a proper limit of a finite \RSB{}
solution with e.g. $k$ \emph{steps}. In the previous sections we have described its
statistical structure, that depends only on the function that we called $y(q)$ (related
to the distribution of overlaps between pairs of states): given a reference node with
energy $\bar{U}_1$, and for all energy intervals $(\bar{U}_2, \bar{U}_2+\dd\bar{U}_2)$,
one adds a branch $\bar{U}_1\rightarrow\bar{U}_2$ with probability $e^{y(q_1)\cdot(\bar{U}_2-\bar{U}_1)}\dd\bar{U}_2$;
in the same way one adds all the branches up to the $k+1$-st level, according to
$\Prob[\mathrm{branch\ }\bar{U}_i\rightarrow(\bar{U}_{i+1},\bar{U}_{i+1}+\dd\bar{U}_{i+1})] = e^{y(q_i)\cdot(\bar{U}_{i+1}-\bar{U}_i)}\dd\bar{U}_{i+1}$.
Note that there is an infinite number of branches at any step; the values appearing
inside the function $y(q)$ are a discretization of the interval $[0,1]$, ${\{q_i\}}_{i=1}^{k}$
(with $q_{k+1}=1$). This process defines a tree, an example of which is shown in
\Fref{fig:appendixultrametrictree} \emph{left}. The (non-extensive part of the) energy of the states are the values
$\bar{U}$ at the level $k+1$: we call them $\{U_\alpha \}$, without a bar in order to
distinguish them from the intermediate nodes that are used only to define the distribution.\\ \\
The stresses in the states of a system of spheres are generated via a diffusion process
on this same tree --- this holds for other systems too, e.g.\ for the magnetizations
in the Sherrington-Kirkpatrick model. In other words, starting from a reference stress
$\bar{\Sigma}_1$ at the root of the tree, the stress of a node $\bar{\Sigma}_{i+1}$
is distributed as $\Prob[\bar{\Sigma}_{i+1}|\bar{\Sigma_i}] = {(4\pi(q_{i+1}-q_i))}^{-\frac12} e^{-\frac{{(\bar{\Sigma}_{i+1}-\bar{\Sigma}_i)}^2}{4(q_{i+1}-q_i)}}$,
$\bar{\Sigma}_i$ being the stress of the ancestor node in the tree. As before, the
process is stopped at the $k$-th level, when states are reached.
At this point, in order to perform the calculation~\eqref{eq:marginal} we use the
fact that the states are arranged with this \emph{ultrametric} structure. It is useful
to introduce a partition of the states as follows: we call $q_i$-\emph{cluster} the
set of all states with overlap $q_i$ with the ground state; the clusters are all disjoint
and conditionally independent, and they include all the states (in particular the
$q_{k+1}$-cluster contains the ground state only since it is the only state with overlap
$q_{k+1}\equiv 1$ with itself). A key-point in our calculation consists in noticing
the recurrent structure of the tree, namely that if $q_j < q_i$, the $q_j$-cluster
is \emph{larger}, in the sense that the common ancestors of all its states is further
back along the tree, and this in some way it contains more branching levels; in particular,
the $q_i$-cluster is a collections of smaller clusters that are statistically equivalent
to the $q_{i+1}$-cluster. Note that at this point the overlaps $q_i$ are both a measure
along the depth of the tree and along its breadth: both concepts are related to the dimension of the $q_i$-clusters.\\ \\
Keeping in mind the structure of the states presented in \Fref{fig:ultrametrictree}
\emph{left}, the first step in the computation of~\eqref{eq:marginal} is writing the
probability distribution of the total energy minimum in each cluster. It is possible
to compute this function via a recurrent procedure, from smaller to larger clusters.
The ground state, being the only state in its $q_{k+1}$-cluster, is the minimum in
the $q_{k+1}$-cluster. Then we compute probability that there is a state in the $q_k$-cluster
with energy $U_k$, and that it has stress $\Sigma_k$; this, conditionally on the ancestor
node having some energy $\bar{U}_k$ and stress $\bar{\Sigma}_k$, is just the product
of the Poisson and Gaussian variables:

\begin{equation}
p_k(U_k,\Sigma_k|\bar{U}_k,\bar{\Sigma}_k)\dd U_k\dd \Sigma_k \equiv \frac{e^{y_k(U_k - \bar{U}_k) - \frac{{(\Sigma_k - \bar{\Sigma}_k)}^2}{2(q_k - q_{k-1})}}}{\sqrt{2\pi(q_k - q_{k-1})}}\dd U_k\dd \Sigma_k.
\end{equation}

\noindent The probability that this is the ground state of the $q_k$-cluster (that is, there are no states in the same cluster with smaller total energy) is the probability that there is no other state with total energy less than $U_k-\delta\gamma\Sigma_k$. Since all the nodes are conditionally independent (conditional on the ancestor nodes),

\begin{eqnarray}
\fl \mu_k(U_k,\Sigma_k|\bar{U}_k,\bar{\Sigma}_k)\dd U_k\dd \Sigma_k \equiv \\
\equiv p_k(U_k,\Sigma_k|\bar{U}_k,\bar{\Sigma}_k)\dd U_k\dd \Sigma_k\ \cdot\ \prod_{\mathclap{\substack{-\infty\leq U,\Sigma\leq\infty \\ U-\delta\gamma\Sigma < U_k-\delta\gamma\Sigma_k}}}\ \left[1 - p_k(U,\Sigma|\bar{U}_k,\bar{\Sigma}_k)\dd U\dd \Sigma\right] =\\
= p_k(U_k,\Sigma_k|\bar{U}_k,\bar{\Sigma}_k)\dd U_k\dd \Sigma_k\ \cdot\ \exp\left[-\frac{c_k}{y_k} e^{y_k\left(U_k-\bar{U}_k - \delta\gamma(\Sigma_k-\bar{\Sigma}_k)\right)}\right],
\end{eqnarray}\\

\noindent where $c_k$ is a constant that can be computed and does not depend on the energies and stresses. It is possible to define analogous functions $\mu_i(U_i,\Sigma_i|\bar{U}_i,\bar\Sigma_i)$ for all the clusters, and compute them using the self-similarity of the tree: a $q_i$-cluster is the disjoint union of smaller clusters, whose statistics is the same as the (smaller) $q_{i+1}$-cluster. Let's call $\mu_{i+1}(U_{i+1},\Sigma_{i+1}|\bar{U}_{i+1},\bar{\Sigma}_{i+1})$ the probability density of the $q_{i+1}$-cluster's ground state. In order to compute $\mu_i(U_i,\Sigma_i|\bar{U}_i,\bar{\Sigma}_i)$, we first count the probability density $\nu_i(U_i,\Sigma_i|\bar{U}_i,\bar{\Sigma}_i)$ of the ground states $(U_i,\Sigma_i)$ in the smaller subclusters:

\begin{multline}
\nu_i(U_i,\Sigma_i|\bar{U}_i,\bar{\Sigma}_i)\dd U_i\dd \Sigma_i \equiv \dd U_i\dd \Sigma_i \times \\
\times\int \mu_{i+1}(U_i,\Sigma_i|\bar{U}_{i+1},\bar{\Sigma}_{i+1}) p_{i+1}(\bar{U}_{i+1},\bar{\Sigma}_{i+1}|\bar{U}_i,\bar{\Sigma}_i) \dd\bar{U}_{i+1}\dd\bar{\Sigma}_{i+1},
\end{multline}\\

\noindent where $p_i$ is the probability density of branches (product of the Poisson and Gaussian distributions):

\begin{equation}
p_i(\bar{U}_i,\bar{\Sigma}_i|\bar{U}_{i-1},\bar{\Sigma}_{i-1})\dd\bar{U}_i\dd\bar{\Sigma}_i \equiv \frac{e^{y_i(\bar{U}_i-\bar{U}_{i-1}) - \frac{{(\bar{\Sigma}_i-\bar{\Sigma}_{i-1})}^2}{2(q_i - q_{i-1})}}}{\sqrt{2\pi(q_i - q_{i-1})}}\dd\bar{U}_i\dd\bar{\Sigma}_i.
\end{equation}\vspace{1em}

\noindent Once $\nu_i$ is computed it is possible to calculate $\mu_i$, again with the formula

\begin{eqnarray}
\fl \mu_i(U_i,\Sigma_i|\bar{U}_i,\bar{\Sigma}_i)\dd U_i\dd \Sigma_i \equiv \\
\equiv \nu_i(U_i,\Sigma_i|\bar{U}_i,\bar{\Sigma}_i)\dd U_i\dd \Sigma_i\ \cdot\ \prod_{\mathclap{\substack{-\infty\leq U,\Sigma\leq\infty \\ U-\delta\gamma\Sigma < U_i-\delta\gamma\Sigma_i}}}\ \left[1 - \nu_i(U,\Sigma|\bar{U}_i,\bar{\Sigma}_i)\dd U\dd \Sigma\right].
\end{eqnarray}\\

\noindent The results for $\nu_i(U_i,\Sigma_i|\bar{U}_i,\bar\Sigma_i),\mu_i(U_i,\Sigma_i|\bar{U}_i,\bar\Sigma_i)$ are:

\begin{eqnarray}
\fl \nu_i(U_i,\Sigma_i|\bar{U}_i,\bar{\Sigma}_i) = d_i \frac{\exp\left[y_i(U_i - \bar{U}_i) - \frac{{(\Sigma_i - \bar{\Sigma}_i)}^2}{2(q_i - q_{i-1})} + \delta\gamma f_i(\Sigma_i - \bar{\Sigma}_i)\right]}{\sqrt{2\pi(q_k - q_{i-1})}},\\
\fl \mu_i(U_i,\Sigma_i|\bar{U}_i,\bar{\Sigma}_i) = \nu_i(U_i,\Sigma_i|\bar{U}_i,\bar{\Sigma}_i) \exp\left[-\frac{c_i}{y_i} e^{y_i\left(U_i - \bar{U}_i - \delta\gamma(\Sigma_i - \bar{\Sigma}_i)\right)}\right],
\end{eqnarray}\\

\noindent where $c_i,\ d_i$ are constants that do not depend on energies and stresses, and $f_i = \frac{\sum_{j=i}^k y_j}{k+1-i}-y_i$. Notice that all these functions depend only on $U_i - \bar{U}_i,\ \Sigma_i - \bar{\Sigma}_i$.\\ \\
\noindent We now compute the joint probability density of all clusters' minima (after subtracting the energies and stresses of the unperturbed ground state):

\begin{multline}
\Prob[\Delta U_{k+1},\Delta\Sigma_{k+1};\Delta U_k,\Delta\Sigma_k;\dots;\Delta U_1,\Delta\Sigma_1| \delta\gamma] \equiv \\
\equiv \int \left[\prod_{i=1}^k \dd \bar{U}_i\dd \bar{\Sigma}_i\right] p_k(U_\gs+\Delta U_{k+1},\Sigma_\gs+\Delta\Sigma_{k+1}|\bar{U}_k,\bar{\Sigma}_k) \left[\prod_{i=1}^{k-1} p_i(\bar{U}_{i+1},\bar{\Sigma}_{i+1}|\bar{U}_i,\bar{\Sigma}_i)\right] \times \\
\times \delta(\Delta U_{k+1})\delta(\Delta\Sigma_{k+1}) \left[\prod_{i=1}^k \mu_i(U_\gs+\Delta U_i,\Sigma_\gs+\Delta\Sigma_i|\bar{U}_i,\bar{\Sigma}_i)\right].
\end{multline}\\

\noindent The variables $U_{k+1},\ \Sigma_{k+1}$ (forced to be 0 by the delta functions) have been introduced to take into account the level crossing with the unperturbed ground state --- we are going to constraint the new ground state to be smaller than the unperturbed one. After some manipulations and change of variables, we are able to cast this probability as

\begin{multline}
\label{eq:jointpdfclusters}
\Prob[\Delta U_{k+1},\Delta\Sigma_{k+1};\Delta U_k,\Delta\Sigma_k;\dots;\Delta U_1,\Delta\Sigma_1| \delta\gamma] = \delta(\Delta U_{k+1})\delta(\Delta\Sigma_{k+1}) \times \\
\times \prod_{i=1}^k \Delta y_i\, e^{-\Delta y_i\, \Delta U_i} \int\dd^k \underline{x}\, K(\underline{x}) \prod_{i=1}^k \frac{e^{-\frac{{(x_i - \Delta\Sigma_i + \delta\gamma z_i)}^2}{2\Delta q(k+1-i)}}}{\sqrt{2\pi\Delta q(k+1-i)}},
\end{multline}\\

\noindent where we have chosen $q_i - q_{i-1} \equiv \Delta q$ for every $i$, without loss of generality; $\Delta y_i \equiv y_i - y_{i-1} \rightarrow y^\prime(q_i)\, \dd q$ ($y_0 \equiv 0$); $1-q_i \equiv \Delta q(k+1-i)$; here $\underline{x} = (x_1,\dots,x_k)$ and $K(\underline{x}) = {(2\pi\Delta q)}^{-\frac{k}{2}}e^{-\frac{x_k^2}{2\Delta q}} \prod_{j=1}^{k-1} e^{-\frac{{(x_{j+1} - x_j)}^2}{2\Delta q}}$ is a Gaussian kernel. The constant $z_i$ is $Y_i + (1-q_i) y^\prime(q_i)\Delta q$, with $Y_i = \Delta q \sum_{j=i}^k y_j \rightarrow Y(q_i)=\int_{q_i}^1\dd q\, y(q)$ in the $k\rightarrow\infty$ limit. Notice also that the dependence on the energies and on the stresses is completely decoupled.\\ \\
Performing $k$ Hubbard-Stratonovich transformations, it is possible to compute the Gaussian integrals explicitly:

\begin{multline}
\Prob[\Delta U_{k+1},\Delta\Sigma_{k+1};\Delta U_k,\Delta\Sigma_k;\dots;\Delta U_1,\Delta\Sigma_1| \delta\gamma] =\\
= \delta(\Delta U_{k+1})\delta(\Delta\Sigma_{k+1}) \left[\prod_{i=1}^k \Delta y_i\, e^{-\Delta y_i\, \Delta U_i}\right] \times \\
\times \int \left[\prod_{i=1}^k \frac{\dd u_i}{2\pi}e^{-\frac12 (1-q_i)u_i^2-iu_i (\Delta\Sigma_i-\delta\gamma z_i)}\right] \int\dd^k \underline{x}\, K(\underline{x}) \prod_{i=1}^k e^{iu_i x_i} =\\
= \delta(\Delta U_{k+1})\delta(\Delta\Sigma_{k+1}) \left[\prod_{i=1}^k \Delta y_i\, e^{-\Delta y_i\, \Delta U_i}\right] \left[\prod_{i=1}^k \int\frac{\dd u_i}{2\pi}e^{-2\cdot\frac12 (1-q_i)u_i^2-iu_i (\Delta\Sigma_i-\delta\gamma z_i)}\right] =\\
= \delta(\Delta U_{k+1})\delta(\Delta\Sigma_{k+1}) \prod_{i=1}^k \Delta y_i\, e^{-\Delta y_i\, \Delta U_i} \frac{e^{-\frac{{(\Delta\Sigma_i-\delta\gamma z_i)}^2}{4(1-q_i)}}}{\sqrt{4\pi(1-q_i)}}.
\label{eq:splitpdfclusters}
\end{multline}\\

\noindent From \Eref{eq:splitpdfclusters} we can extract the probability distribution $\Prob[\Delta U_j,\Delta\Sigma_j,j|\delta\gamma]$ that the new ground state lies in the $q_j$-cluster and that it has energy $U_\gs+\Delta U_j$ and stress $\Sigma_\gs+\Delta\Sigma_j$:

\begin{multline}
\Prob[\Delta U_j,\Delta\Sigma_j,j|\delta\gamma] = \int\left[\prod_{i=1,\ i\neq j}^{k+1} \dd \Delta U_i\dd \Delta\Sigma_i\, \theta(\Delta U_i-\delta\gamma\Delta\Sigma_i>\Delta U_j-\delta\gamma\Delta\Sigma_j)\right]\times \\
\times \Prob[\Delta U_{k+1},\Delta\Sigma_{k+1};\Delta U_k,\Delta\Sigma_k;\dots;\Delta U_1,\Delta\Sigma_1|\delta\gamma] =\\
= \delta(k+1-j)\delta(\Delta U_j)\delta(\Delta \Sigma_j) \prod_{i=1}^k \chi_i(0)\ +\ \hat\theta(k+1-j)\theta(\Delta U_j-\delta\gamma\Delta\Sigma_j<0)\times \\
\times \frac{\Delta y_j}{\sqrt{4\pi(1-q_j)}} e^{-\Delta y_j\, \Delta U_j - \frac{{(\Delta\Sigma_j-\delta\gamma z_j)}^2}{4(1-q_j)}} \prod_{i=1,\ i\neq j}^k \chi_i(\Delta U_j-\delta\gamma\Delta\Sigma_j),\label{eq:prefinal}
\end{multline}\\

\noindent where $\hat\theta(k+1-j) \equiv 1$ if $j = 1,\dots,k$ and 0 otherwise, and
\begin{multline}
\chi_i(S,\delta\gamma) = H\left(\abs{\delta\gamma}\, \sqrt{1-q_i}\Delta y_i + \frac{\abs{\delta\gamma}\, Y_i + S/\abs{\delta\gamma}}{\sqrt{1-q_i}}\right) +\\
+ H\left(\abs{\delta\gamma}\, \sqrt{1-q_i}\Delta y_i - \frac{\abs{\delta\gamma}\, Y_i + S/\abs{\delta\gamma}}{\sqrt{1-q_i}}\right) \exp\left \{-\Delta y_i \left(\delta\gamma^2\, Y_i + S\right)\right \}
\end{multline}\\

\noindent with $H(x) \equiv \frac12\, \mathrm{erfc}\!\left(\frac{x}{2}\right) = \frac{1}{\sqrt{\pi}}\int_{x/2}^\infty e^{-t^2}\dd t$. It is now possible to compute the probability distribution for the total energy jumps, by integrating over all the variables with the constraint $\Delta E = \Delta U_j - \delta\gamma\Delta\Sigma_j$:

\begin{multline}
\Prob[\Delta E|\delta\gamma] = \sum_j \int_0^\infty \dd\Delta U \int_{-\infty}^\infty \dd\Delta\Sigma \delta\left(\Delta E - \Delta U + \delta\gamma\Delta\Sigma\right) \Prob[\Delta U,\Delta\Sigma,j|\delta\gamma] =\\
= \delta(\Delta E)\mathcal{R}(0,\delta\gamma) - \theta(-\Delta E)\frac{\partial\mathcal{R}(\Delta E,\delta\gamma)}{\partial\Delta E},
\end{multline}\\

\noindent where $\mathcal{R}(\Delta E,\delta\gamma) \equiv \prod_i \chi_i(\Delta E, \delta\gamma)$.
Taking the limits $k\rightarrow\infty,\, \Delta q \rightarrow 0,\, k\Delta q\rightarrow 1$
and setting $\delta\gamma = \delta\gamma \sqrt{N}$
we find in the end \Eref{eq:final}.\\ \\

\clearpage
\section*{References}
\bibliographystyle{unsrt}
\bibliography{biblio}

\end{document}